\documentclass[final,2p,times,10pt,twocolumn]{elsarticle}

\usepackage{amssymb}
\usepackage[T1]{fontenc}
\usepackage{booktabs}
\usepackage{graphicx}
\usepackage{color}
\usepackage{hyperref}
\usepackage{amssymb,amsmath,bm}
\usepackage{textcomp}
\usepackage{graphicx}
\usepackage{wrapfig}
\usepackage{lipsum}
\usepackage{dblfloatfix}
\usepackage{fixltx2e}
\usepackage{nameref}
\usepackage{amsmath}
\usepackage{hyperref}
\usepackage{graphicx}
\usepackage{varioref}
\usepackage{booktabs}  % nice looking tables
\usepackage{siunitx}
\usepackage{numprint}
\usepackage{textcomp}

\usepackage{lipsum}
\journal{Computer Speech \& Language}
\usepackage{lipsum}
\makeatletter
\def\ps@pprintTitle{%
	\let\@oddhead\@empty
	\let\@evenhead\@empty
	\def\@oddfoot{}%
	\let\@evenfoot\@oddfoot}
\makeatother
\begin{document}
\begin{frontmatter}
%%%% Title, authors and addresses
%%
\title{Using Speech Technology for Quantifying Behavioral Characteristics in Peer-Led Team Learning Sessions}
\author{Harishchandra Dubey, Abhijeet Sangwan\textsuperscript{}, and John H. L. Hansen\tnoteref{t2}\tnotetext[t2]{\textcolor{blue}{This material is presented to ensure timely dissemination of scholarly and technical work. Copyright and all rights therein are retained by the authors or by the respective copyright holders. The original citation of this paper is:  Harishchandra Dubey, Abhijeet Sangwan, John H.L. Hansen, Using speech technology for quantifying behavioral characteristics in peer-led team learning sessions, Computer Speech \& Language, Available online 14 April 2017, ISSN 0885-2308, http://doi.org/10.1016/j.csl.2017.04.002.
\(http://www.sciencedirect.com/science/article/pii/S0885230816303333\).}}}
\address{Center for Robust Speech Systems, Erik Jonsson School of Engineering and Computer Science\\
	The University of Texas at Dallas, Richardson, TX 75080, USA \\
	{\small \tt \{harishchandra.dubey, abhijeet.sangwan, john.hansen\}@utdallas.edu}
}
\begin{abstract}
Peer-Led Team Learning (PLTL) is a learning methodology where a peer-leader co-ordinate a small-group of students to collaboratively solve technical problems. PLTL have been adopted for various science, engineering, technology and maths courses in several US universities. This paper proposed and evaluated a speech system for behavioral analysis of PLTL groups. It could help in identifying the best practices for PLTL. The CRSS-PLTL corpus was used for evaluation of developed algorithms. In this paper, we developed a robust speech activity detection (SAD) by fusing the outputs of a DNN-based pitch extractor and an unsupervised SAD based on voicing measures. Robust speaker diarization system consisted of bottleneck features (from stacked autoencoder) and informed HMM-based joint segmentation and clustering system. Behavioral characteristics such as participation, dominance, emphasis, curiosity and engagement were extracted by acoustic analyses of speech segments belonging to all students. We proposed a novel method for detecting question inflection and performed equal error rate analysis on PLTL corpus. In addition, a robust approach for detecting emphasized speech regions was also proposed. Further, we performed exploratory data analysis for understanding the distortion present in CRSS-PLTL corpus as it was collected in naturalistic scenario. The ground-truth Likert scale ratings were used for capturing the team dynamics in terms of student's responses to a variety of evaluation questions. Results suggested the applicability of proposed system for behavioral analysis of small-group conversations such as PLTL, work-place meetings~\emph{etc.}.
\end{abstract}
\begin{keyword}
Behavioral Speech Processing \sep Bottleneck Features \sep Curiosity \sep Deep Neural Network \sep Dominance \sep Auto-encoder \sep Emphasis \sep Engagement \sep Peer-Led Team Learning \sep Speaker Diarization \sep Small-group Conversations
%% or \MSC[2008] code \sep code (2000 is the default)
\end{keyword}
\end{frontmatter}
%\linenumbers
%% main text
\section{Introduction}
\label{sec:intro}
Peer-Led Team Learning (PLTL) is a structured methodology where a team leader facilitate collaborative problem solving among small-group of students. PLTL have shown positive outcomes towards learning~\citet{snyder2016peer}. The traditional teaching model lacks one-to-one interaction and peer-feedback unlike PLTL. Peer leaders are expected to give helpful hints and comments during students' discussion. Peer leaders are not supposed to reveal solutions, in contrast to the traditional teaching model~\citet{cracolice2001peer}.

We established the CRSS-PLTL corpus in~\citet{dubey2016interspeech} for audio-based analysis of PLTL sessions. Earlier, we developed a robust diarization system that combined bottleneck features (from a stacked autoencoder) with an informed HMM-based joint segmentation and clustering approach~\citet{dubey2016robust}. The minimum-duration of short conversational-turns and number of students were incorporated as side information to the HMM-based diarization system. The output probability density function of each HMM state was modeled using a Gaussian Mixture Model (GMM). Each HMM state was allowed to have several sub-states for ensuring the minimum-duration constraint of conversational-turns. A modified form of Bayesian Information Criterion (BIC) was used for iterative merging and re-segmentation. We continued the merging of HMM states till the number of states was same as that of the speakers. 

\begin{figure*}[!t]
\centering
\includegraphics[width=450pt]{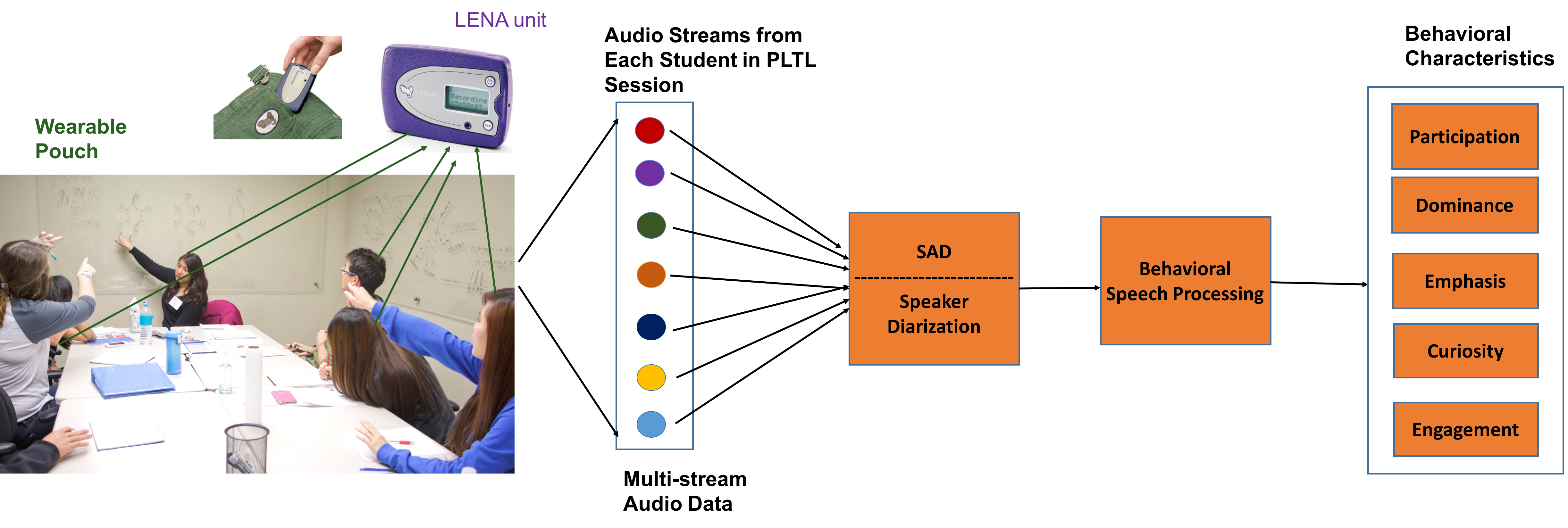}
%\vspace{-2mm}
\caption{Proposed speech system for extracting behavioral characteristics from PLTL sessions. The front-end consists of speech activity detection (SAD) and speaker diarization. It is followed by behavioral speech processing that involved acoustic analyses of individual speaker segments for extraction of behavioral metrics such as participation, dominance, emphasis, curiosity and engagement.}
\label{fig_pipe}
\end{figure*}

Authors established the domain of behavioral signal processing in~\citet{narayanan2013behavioral}. It refers to the use of computational methods and signal processing tools for extracting the behavioral patterns in human-human and human-machine communication. The present paper is a step towards extracting behavioral characteristics of students attending a PLTL session using acoustic analysis of their speech signal. Behavioral speech processing block performed acoustic analyses for extracting  features that encapsulate behavioral aspects of conversations (See Figure~\ref{fig_pipe}). Particularly, the proposed extracted five features namely (1) participation; (2) dominance; (3) emphasis; (4) curiosity and (5) engagement from the speech signal. These features could be used for quantifying the behavioral characteristics in peer-led team learning sessions. 

This paper made the following contributions in area of speech technology for behavioral analysis of PLTL sessions:
\begin{itemize}
\item Improved speech activity detection using DNN-based pitch and TO-comboSAD~\citet{ziaei2014speech}; 
%\item Explorator Data analysis of CRSS-PLTL corpus using ground-truth 
\item Informed-HMM diarization system using bottleneck features obtained from stacked autoencoder; 
\item Extracting behavioral characteristics such as participation, dominance, emphasis, curiosity and engagement features from speech signal;
%and speaker matrix to understand the influence in PLTL discussions
%\item Extracting dominance as individual characteristics
% \item Detecting emphasized speech for identification of 'hot-spots' in a PLTL session where important discussions are located ;
% \item Predicting curiosity in terms of question-inflection detection
% %\item Cohesion as Group characteristics
% \item Measuring engagement using speech rate (word count)
%\item Question Inflection Detection
\end{itemize}
The developed methods were evaluated over disjoint evaluation datasets taken from CRSS-PLTL corpus (See Table~\ref{table_eval_sets}).
\begin{table*}[!t]
\centering
\caption{Description of evaluation datasets derived from CRSS-PLTL corpus that were used for validating the proposed algorithms. The evaluation datasets were disjoint, i.e., chosen from different PLTL session to avoid bias in annotation process.}
\begin{tabular}{|c|c|c|}
\hline
\textbf{Eval dataset}&~\textbf{Duration(minute)}&\textbf{Description} \\\hline
Eval-Set-1 &70& Diarization   \\\hline
Eval-Set-2 &21& Participation \\\hline
Eval-Set-3 &70 & Dominance rating\\\hline
Eval-Set-4 &30& Emphasis \\\hline
Eval-Set-5 &30& Curiosity (question-inflection) \\ \hline
Eval-Set-6 &70& Engagement (speech rate)\\ \hline
Eval-Set-7 &87& Speech Activity Detection\\ \hline
\end{tabular}
\label{table_eval_sets}
\end{table*}
\begin{figure*}[!t]
\centering
\includegraphics[width=450bp]{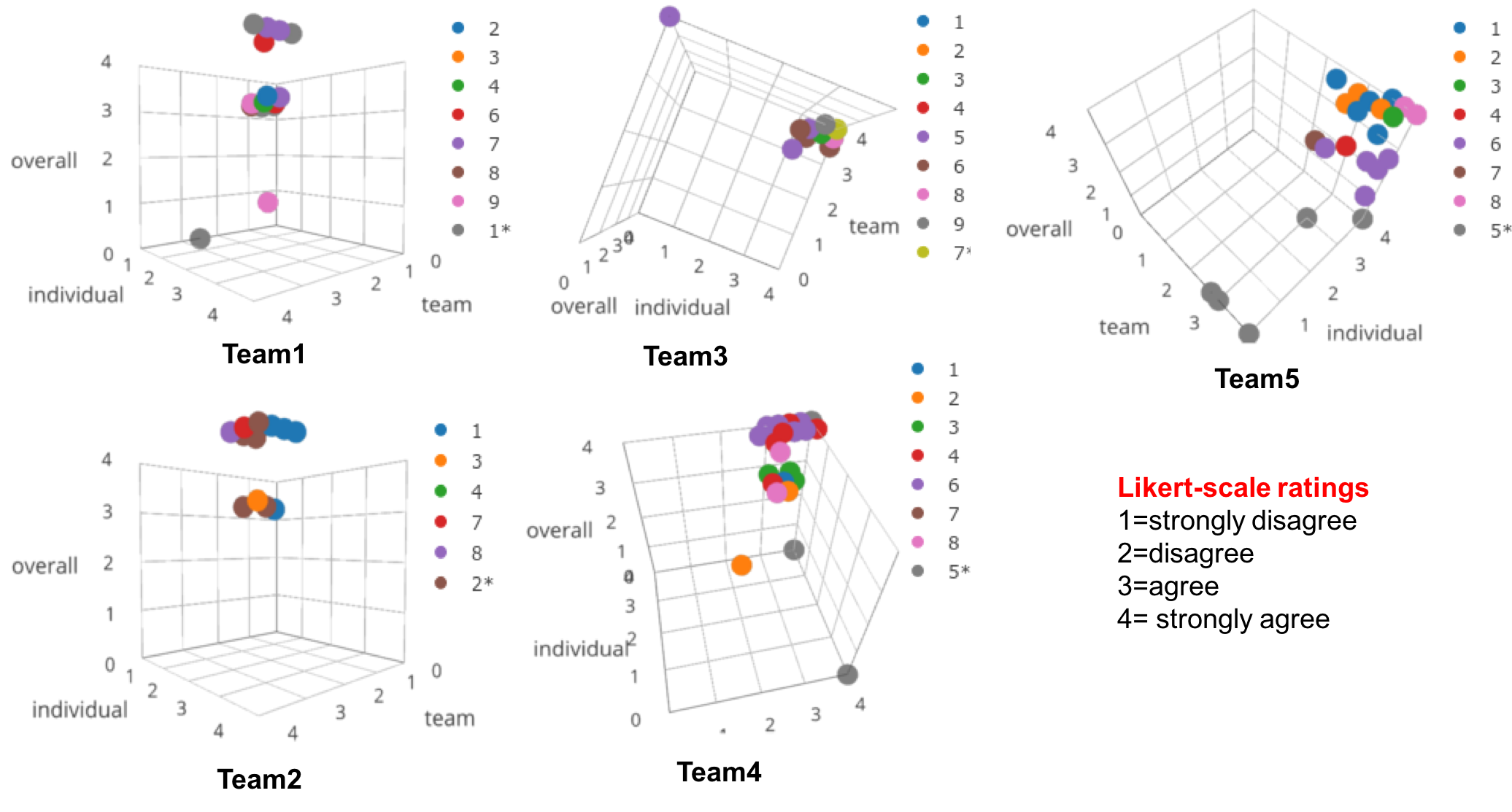}
\caption{Showing overall dynamics of five PLTL teams tracked over eleven weeks in terms of ground-truth Likert-scale ratings obtained from students. These ratings were obtained from feedback forms filled by students after each PLTL session. We discuss more details in Section~\ref{sec:annotation}.}
\label{fig_likert}
\end{figure*}
\begin{table*}[!t]
\caption{The questions designed to assess the ground-truth Likert-scale ratings from students. PLTL group (PG) and students performance (SP) refers to two categories of questions developed to assess the student's view on group characteristics and his/her own characteristics, respectively. The Q8 refers to overall assessment.}
\centering
%\begin{center}
\begin{tabular}{|c|c|c|}
\hline
S.No. & Description & Assessment Type\\
\hline
Q1&My PLTL group was~\textit{friendly} today&PG\\
\hline
Q2&My PLTL group was~\textit{engaging} today&PG\\
\hline
Q3&My PLTL group was~\textit{helpful} today&PG\\
\hline
Q4&My PLTL group was~\textit{motivated} today&PG\\
\hline
Q5&I~\textit{learned} a lot in today's PLTL session&SP\\
\hline
Q6&I felt~\textit{comfortable} with the interaction with my PLTL group today&SP\\
\hline
Q7&My participation in today's PLTL session increased my~\textit{confidence} in the course&SP\\
\hline
Q8&Overall, the PLTL sessions are helping me do better in my course&Overall\\
\hline
\end{tabular}
\label{table_pltl_ques}
%\end{center}
\end{table*}
\begin{table*}[!t]
\caption{Spearman's rank correlation between ground-truth responses of question shown in Table~\ref{table_pltl_ques} for five PLTL groups over 11 sessions for each group, i.e., 55 PLTL sessions in total. We can see high pair-wise correlation in these responses providing hints for combining these into three dimensional scores as shown in Figure~\ref{fig_likert}. We combine PG questions (Q1-Q4) together and SP questions (Q5-Q7) together and left Q8 (overall) as it is. This resulted in three dimensional space for each team that is visualized in Figure~\ref{fig_likert}. The students along with peer leaders are color coded. The peer leaders for each team are marked with asterisk above their numerical index.}
\label{table_pltl_gt_corr}
\centering
%\begin{center}
\begin{tabular}{|c|c|c|c|c|c|c|c|c|}
\hline
 S.No. & Q1 & Q2 & Q3 & Q4 & Q5 & Q6 & Q7 & Q8 \\
\hline
Q1 & 1 & 0.93 & 0.91  & 0.90 & 0.89  & 0.93 & 0.89  & 0.91 \\
\hline
Q2 & 0.93  & 1  & 0.92  & 0.93 &  0.90  & 0.92  & 0.0.90 & 0.89 \\
\hline
Q3 & 0.91 & 0.0.93 & 1 & 0.90 & 0.91 & 0.91 & 0.91 & 0.92 \\
\hline
Q4 & 0.90 & 0.93 & 0.90 & 1 & 0.88  & 0.91 & 0.88 & 0.90\\
\hline
Q5 & 0.89 & 0.90 & 0.92 & 0.88 & 1 & 0.91 & 0.89 & 0.91\\
\hline
Q6 & 0.93 & 0.92 & 0.91 & 0.91 & 0.91 & 1 &  0.92 & 0.92\\
\hline
Q7 & 0.89 & 0.90 & 0.91 & 0.88 & 0.89 & 0.92  & 1 & 0.94 \\
\hline
Q8 & 0.91 & 0.89 & 0.92 & 0.90 & 0.91  & 0.92 & 0.94 & 1\\
\hline
\end{tabular}
\label{table_spearman_corr}
%\end{center}
\end{table*}
\section{Peer-Led Team Learning}
\label{sec:pltl}
Peer-led team learning (PLTL) is a methodology used for improving learning outcomes in small-group of students attending the same course. PLTL had been adopted in several US universities for various undergraduate courses. Each team is assigned a peer leader who coordinated discussions among students, and facilitated collaborative problem solving. The peer leaders had passed the same course in earlier semester and thus they were aware of the challenges in learning the subject. Peer leader knew the strategies that could help in mastering the technical content of the course. Peer leaders were not supposed to tell the solutions, rather they provided helpful hints and direction that could guide the students to collaboratively solve the problems.
\subsection{CRSS-PLTL Corpus}
\label{sec:corpus}
This section briefly describes the CRSS-PLTL corpus that motivated the research discussed in this paper. The CRSS-PLTL corpus contains audio data from five PLTL teams for eleven weeks. Thus, a total of 55 PLTL sessions were recorded, each of which have multi-stream audio data. Each PLTL team had five to eight students and a peer leader. All the five PLTL teams were attending an undergraduate Chemistry course at the University of Texas at Dallas. The audio data collection was longitudinal over a three month window. Each PLTL session was organized for approximately 70 to 80 minutes. Each student wore a wearable pouch containing LENA digital recorder as shown in Figure~\ref{fig_pipe}. Thus, we have as many audio streams as students attending the PLTL session. LENA device could record audio signals for long-duration of up to sixteen hours and had been used for a variety of human-to-human communication research, for example adult-child interaction~\citet{sangwan2015studying}~\emph{etc.}. 

\begin{figure}[!t]
\centering
\includegraphics[width=240pt]{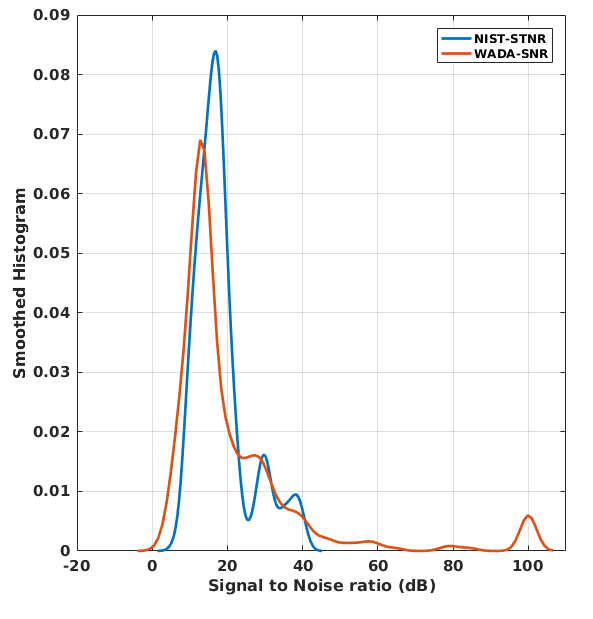}
\caption{The distribution of WADA-SNR~\citet{kim2008robust} and NIST-STNR~\citet{niststnr} signal to noise ratios(SNRs). Five-minute segments were processed to generated these ratios. We used total of three teams with nine channels each. All teams participated in 80 minute PLTL session, so in total 36 hours of data was used for generating this figure. Since all PLTL sessions were carried out in same space, we could not observe any significant difference in this plot by using more data. NIST STNR have tri-modal distribution while WADA-SNR had bi-modal distributions. We can see that the majority of the segments have SNR between 0 to 15 dB that shows moderate to high noise levels in PLTL data. In addition, huge reverberation is another challenge.}
\label{fig_snr}
\end{figure}

The CRSS-PLTL data contained huge reverberation and noise that pose challenges for speech processing~\ref{fig_snr}. Several instances of significant degradation due to noise and reverberation were common. During such instances, the speech was not intelligible over several streams. After the PLTL session was concluded, each student and the peer leader completed a questionnaire. The questionnaire sought Likert-scale ratings for subjective questions regarding behavior, communication and learning~\emph{etc.} as given in Table~\ref{table_pltl_ques}. We would discuss these questions in Section~\ref{sec:annotation}, and Table~\ref{table_spearman_corr}. The ground-truth Likert scale ratings were combined into three dimensional scores as explained in Section~\ref{sec:annotation} and visualized in Figure~\ref{fig_likert}. Earlier, we introduced the CRSS-PLTL corpus in~\citet{dubey2016interspeech}.
\subsection{Listening Tests~\& Annotation}
\label{sec:annotation}
Table~\ref{table_eval_sets} shows the duration and brief description of six evaluation set that were derived from CRSS-PLTL corpus for validation experiments. These seven evaluation datasets were annotated for diarization segments (Eval-Set-1), speech activity detection (Eval-Set-7), participation (Eval-Set-2), dominance(Eval-Set-3), emphasis (Eval-Set-4), curiosity in terms of question-inflection (Eval-Set-5), and engagement in terms of speech rate (Eval-Set-6). The duration of these evaluation sets were provided in Table~\ref{table_eval_sets}. For analysis of behavioral characteristics such as dominance, emphasis, curiosity (in terms of question-inflection), engagement, we performed intelligent listening test. The annotators performed the listening test for labeling the behavioral characteristics. Different evaluation sets were made out of CRSS-PLTL data for evaluation of each of the behavioral characteristics. Using different (disjoint) evaluation dataset was to make sure that annotation bias was the least as some of the studied behavioral characteristics were correlated. 

The evaluation set for question-inflection detection was thirty minute audio data. We used another thirty minutes of data for evaluation of emphasis detection algorithm. For the listening test, an annotator marked the start-time and end-time of audio segments composed of (1) emphasized-speech regions and (2) interrogative utterances/questions. The goal of audio analysis was to detect the temporal boundaries of segments with emphasized-speech and question-inflection. 

It is important to note that the semantic aspects were taken into account during ground-truth annotations. For instance, the emphasis was marked on the basis of what was said and how it was spoken in the given context. The same procedure was applied for question-inflections. However, the algorithms developed for detecting these two phenomenon are based on acoustics features only (fundamental frequency and energy). The speaker diarization ground-truth was obtained on Eval-Set-1 that contained 70 minute audio. Participation refers to annotating the percentage time for which a speaker was active in the PLTL session (Eval-Set-2). For measuring the engagement in terms of speech rate, annotators listened to each five minute segment of PLTL session and note down the number of words spoken. Five minutes segment were derived for Eval-Set-6 that consists of 70 minutes audio data. We discuss the results in Section~\ref{sec:results}. The speech activity detection was evaluated on 87 minutes of audio data from a PLTL session (Eval-Set-7). 

The dominance ratings (ground-truth) were obtained on each five minute segment of Eval-Set-3 (70 minutes). There were seven students in Eval-Set-3. For each five-minute segment, we compute a dominance score ($DS$) for each of the seven students using unsupervised acoustic analysis explained in Section~\ref{sec:dominance}. Each five-minute segment of Eval-Set-3 was assigning a ground-truth dominance rating ($D_{rate}$) for each student per segment. Three annotators listened to each five-minute segment and assigned a dominance rating ($D_{rate}$) for each student per segment. The ground-truth dominance rating, $D_{rate}$, was a number between 1 and 5. The speakers who were present in the whole session but did not speak in the chosen segment were assigned a dominance rating, $D_{rate}=1$. The scores of $D_{rate}=2$ and $D_{rate}=5$ were assigned to the least-and most-dominant students who spoked in that segment. For students who spoke in that segment and were neither least-dominant nor most-dominant, we assigned them a $D_{rate}$ between 2.25 and 4.75. It was possible to score 2.25, 2.50, 2.75, 3.0, 3.25, 3.50, 3.75, 4.0, 4.25, 4.50 and 4.75. However, no fractions other than these were used to ensure consistency in evaluations. We averaged the ground-truth rating ($D_{rate}$) of all three annotator to get a final ground-truth that was used for computing the correlation with unsupervised dominance score ($DS$).
\subsection{Exploratory Data Analysis}
\label{sec:data-analysis}
In this section, we discuss general characteristic of PLTL data. Figure~\ref{fig_snr} shows the distribution of WADA-SNR~\citet{kim2008robust} and NIST-STNR~\citealt{niststnr} signal to noise ratios computed over five-minute segments of 36 hours of PLTL data from three teams. Each team has nine audio streams. The NIST-STNR has tri-modal distribution with significant first model. One the other had, WADA-SNR has bi-modal distribution where first model is significant. The SNR over five-minute segments was mostly between 0 and 15 dB that showed moderate-to-high noise levels. In addition, huge reverberation was also present that could not be visualized in this plot.

At the end of PLTL sessions, each student and their peer leader completed a form that contained eight behavioral question with four options on Likert-scale (see Figure~\ref{fig_likert}). Questions (Q1, Q2,...,Q8) were given in Table~\ref{table_pltl_ques}. These questions belong to three categories, namely PLTL group (PG) assessment, students performance (SP) and overall. The questions Q1, Q2, Q3 and Q4 were regarding the PLTL group (PG) and questions Q5, Q6 and Q7 were based on students performance (SP). The last question, Q8 summarizes the overall assessment. These responses were done on a Likert-scale with four choices, namely strongly disagree(1), disagree(2), agree(3), strongly agree(4) (see Figure~\ref{fig_likert}). Each of these eight questions had a response from each student while team leader responded to only PG and overall category of questions.

Table~\ref{table_pltl_ques} shows the statement of these questions and its categorization as PLTL group (PG) assessment, students performance (SP) and overall. The Spearman's rank correlation uses ranks instead of the actual values used by the Pearson's correlation. Table~\ref{table_pltl_gt_corr} shows the pair-wise Spearman's rank correlation between ground-truth responses of each question. We could see that among pair-wise correlation between questions Q1 to Q4, the minimum and maximum values were 90\% and 93\% respectively. The same values for questions Q5, Q6 and Q7 were 89\% and 92\%. This showed the responses were consistent with respect to categorization. If we see the correlation between Q8 and other questions, we have minimum and maximum values of 89\% and 94\%. This table gave hints that instead of using responses from eight questions, we could reduce this to a smaller set.

Finally, we averaged the responses to question Q1 to Q4 and called it $team$ feature. Similarly, the average of Q5, Q6 and Q7 was called as $individual$ feature. The Q8's response was denoted as $overall$ feature. We did the averaging operations over all responses from each participant. Figure~\ref{fig_likert} showed these three features $team$, $individual$, and $overall$ for all sessions of each team separately. This serves as visualization of behavioral dynamics of each team. 

\begin{figure}[!t]
\centering
\includegraphics[width=230pt]{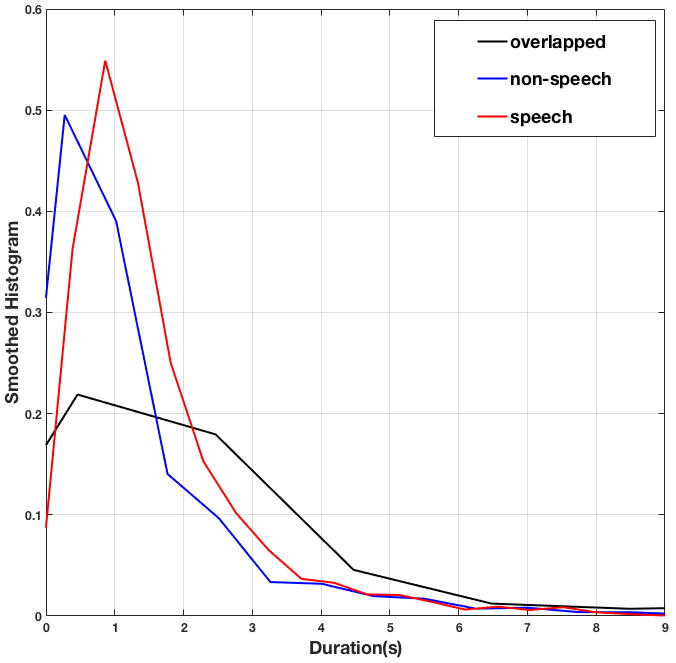}
\caption{Showing distribution of duration of segments with speech, non-speech and overlapped-speech. We could see that most of the segments had duration less than one second. Short-duration segments posed challenge in speaker diarization and behavioral speech processing. The overlapped speech and non-speech accounted for 28.71\% and 29.57\% of total duration leaving behind only 41.72\% speech.}
\label{fig_turn_hist}
\end{figure}

Figure~\ref{fig_turn_hist} showed the distribution of duration of segments with speech, non-speech and overlapped-speech. We could see that most segments were short with less than 1 second duration. Short-duration segments were challenging with respect to speaker diarization and behavioral speech processing. We used ground-truth information from a PLTL session with approximately 87 minutes of audio data for generating this figure (Eval-Set-7, see Table~\ref{table_eval_sets}). The overlapped-speech and non-speech accounted for 28.71\% and 29.57\% of total duration leaving behind 41.72\% speech. The total number of overlapped-speech, non-speech and speech segments were 205, 738 and 1316 respectively, for this data. The data used for this analysis were Eval-Set-7 as given in Table~\ref{table_eval_sets}. We used this dataset for validating the speech activity detection based on fusion of DNN-based pitch estimation and TO-combo-SAD~\citet{sadjadi2013unsupervised,ziaei2014speech}. The results were shown in Table~\ref{table_sad}. 

\begin{figure}[!t]
\centering
\includegraphics[width=240pt]{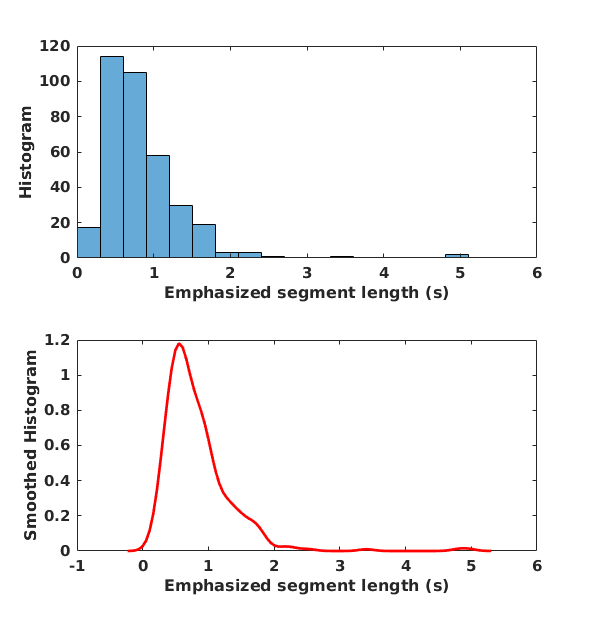}
%\vspace{-2mm}
\caption{Showing distribution of emphasized segment duration for a PLTL session that consisted of approximately 80 minutes audio data. Eight student participated in this session. We could see that most of the emphasized segments have duration less than 1 second.}
\label{fig_emp_hist}
\end{figure}

Figure~\ref{fig_emp_hist} shows the distribution of duration of emphasized speech segments. Ground-truth information from a PLTL session with approximately 80 minute duration was used for generating this figure. It showed that most of the emphasized segments had duration less than 1 second.
\section{Proposed Speech System}
\label{sec:proposed}
Figure~\ref{fig_pipe} shows the block diagram of the proposed speech pipeline. The multi-stream audio data from PLTL sessions was first processed with speech activity detection (SAD) and speaker diarization module to get the individual speaker segments. The output of diarization system facilitate behavioral speech processing in second stage. The behavioral speech processing refers to a set of acoustic analyses that extracts the behavioral characteristics such as participation, dominance, emphasis, curiosity and engagement from speech segments. 
\subsection{Speech Activity Detection}
\label{sec:sad}
Speech activity detection (SAD) was evaluated on Eval-Set-7 (See Table~\ref{table_eval_sets} and the dataset was explained in Section~\ref{sec:data-analysis}. The evaluation results of SAD algorithms are collected in Table~\ref{table_sad}). Figure~\ref{fig_turn_hist} shows the distribution of duration of speech, non-speech and overlapped-speech segments. Non-speech often contained several noise sources such as mumbling of far-speakers, writing-on-the-white-board noise (impulsive) in addition to noise from fan and other background sources. 

We used DNN-based pitch extractor(see Section~\ref{sec:pitch}) along with TO-combo-SAD~\citet{ziaei2014speech} for SAD. The frames that were assigned zero (0 Hz) pitch were declared non-speech. TO-combo-SAD~\citet{sadjadi2013unsupervised, ziaei2014speech} was SAD system developed for DARPA RATS data. TO-combo-SAD had shown good performance on naturalistic audio streams such as NASA Apollo mission data. TO-combo-SAD assigned zero (0) for non-speech and one (1) for speech. We fused the output of both systems for accurate speech activity detection. The frames with non-zero pitch were taken as speech frames and assigned one (1) as SAD output. If both system's output (DNN-based pitch and TO-combo-SAD) were not same, we consider those frames as non-speech. As a results, false alarms were greatly reduced. The non-speech in evaluation dataset has multiple simultaneous sources that results in high false alarm for individual SAD system. We evaluated the SAD system on Eval-Set-7 data as shown in Table~\ref{table_sad}. $Pmiss$ and $Pfa$ refers to miss rate (true-speech detected as non-speech in $\%$) and false alarm rate (true non-speech detected as speech in $\%$), respectively. 

In addition to the proposed fused SAD system, we used a supervised SAD system trained on DARPA RATS data~\citet{van2013robust} and compare its performance with proposed SAD system. The comparison results were shown in Figure~\ref{table_sad}. This was a supervised Neural Network-based SAD system. The Gammatone, Gabor, long-term spectral variability and voicing features were combined together and used for training the neural network. This system was developed for DARPA Robust Automated Transcription of Speech (RATS) program~\citet{van2013robust}. Authors extracted features using speech characteristics such as spectral shape, spectro-temporal modulations, periodicity (pitch harmonics), and long-term spectral variability. Authors used the features from long context-windows to get combined feature vector. These features were used for training a neural network~\citet{van2013robust}. The evaluation on DARPA RATS corpora showed accurate results, thus validating the applicability of
developed SAD system for highly distorted conditions such as those in DARPA RATS~\citet{van2013robust}.

It is important to note that the PLTL data has (1) not-so-close microphone; and (2) small movement in students, such as moving to white board and writing something, was frequent event that made SAD a challenging task. In addition, huge reverberation and noise corrupted the speech data further. The informed HMM-based diarization system was shown in Figure~\ref{fig_diar}. We would discuss the proposed bottleneck features and informed HMM-based diarization system in Section~\ref{sec:diar}. 
\begin{table*}[!t]
\centering
\caption{System parameters for robust pitch extraction method as depicted in Figure~\ref{fig_pitch_blk}. The pitch was used for measuring curiosity (in terms of question-inflection) and emphasis detection. The super-segments of size 2s were used for detecting emphasis and question-inflection.}
\begin{tabular}{|c|c|}
\hline
\textbf{Parameter} & \textbf{Value}\\\hline
Sampling rate &8000Hz\\ \hline
Frame rate & 25ms \\ \hline
Skip-rate & 10ms \\ \hline
Super-segment size & 2s \\ \hline
%Super-segment skip-rate & 1s \\ \hline
Features & Pitch Estimation Filter with Amplitude
Compression~\citet{gonzalez2014pefac}\\ \hline
Splicing context (past) & 2 frames\\\hline
Splicing context (future) & 2 frames\\ \hline
Number of Hidden Layers in DNN&3\\ \hline  
Number of Hidden Nodes (three layers)&1600\\ \hline  
Hidden Layer activation& Sigmoid\\ \hline  
Output Layer activation& Soft-max\\ \hline  
Output Layer dimension (pitch states)&68\\ \hline  
\end{tabular}
\label{table_params_pitch}
\end{table*}
\subsection{Speech Energy}
\label{sec:wpd}
Earlier, we used the formant energy for computing the speaker energy. This energy was leveraged for separating the primary and secondary speakers on each channel of the multi-channel PLTL data (wearer was primary speaker and rest secondary)~\citet{dubey2016interspeech}. More often than not, the wearer was assumed to be the closest to their LENA device as compared to other LENA devices. Thus, the audio channel with highest energy could identify the primary speaker. These intensity differences helped in refining diarization output in one of our previous studies~\citet{dubey2016interspeech}. Later, we computed the energy of speech signal using wavelet packet decomposition~\citet{dubey2016robust}. We choose wavelet packets over formant energy that was used in our earlier studies~\citet{dubey2016interspeech}. 

Formant energy was noise-robust, unlike short-time Fourier transform at the cost of huge computational load. Wavelet packet decomposition was noise-robust and possessed good resolution in time-frequency space with moderate computational load~\citet{wickerhauser1991lectures}. Wavelet packets provided good time-frequency resolution with reasonable computational expense. The position, scale and frequency parameters characterize the wavelet packets~\citet{wickerhauser1991lectures}. Traditional wavelet decomposition had only two parameters, namely (1) position; and (2) scale. Wavelet packets could be viewed as a generalized form of wavelet decomposition. Wavelet packets provide better signal resolution in terms of scale, position and frequency dependence. Wavelet packets are bases generated from decomposition of a signal using orthogonal wavelet functions. There are several computationally simple methods for estimating wavelet packets, that made them a better choice for signal decomposition than computationally expensive continuous wavelet transforms.

Traditional wavelet decomposition generates approximation coefficient vector and detailed coefficient vector after first level of decomposition. At next level and successive levels of decomposition, only approximation coefficient vector is re-decomposed into its approximate and detailed components. On the other hand, wavelet packet decomposition allows each detailed coefficient vector to be decomposed in the same way as the approximate coefficient vector~\citet{wickerhauser1991lectures}. For a speech segment, wavelet packet decomposition generated a complete binary tree allowing a more generic decomposition of the signal. Symlets6 (sym6) wavelet with six levels of decomposition were used for computing the energy. We added the squared wavelet packet coefficient corresponding to the frequency range [50, 2000] Hz for capturing the speech intensity while ignoring the spurious background artifacts and noise. We used the speaker energy for estimating the unsupervised dominance score as discussed in Section~\ref{sec:dominance} and also in emphasis detection (see Section~\ref{sec:emphasis}).
\subsection{Robust Pitch Estimation}
\label{sec:pitch}
\begin{figure}[!t]
\centering
\includegraphics[width=230pt]{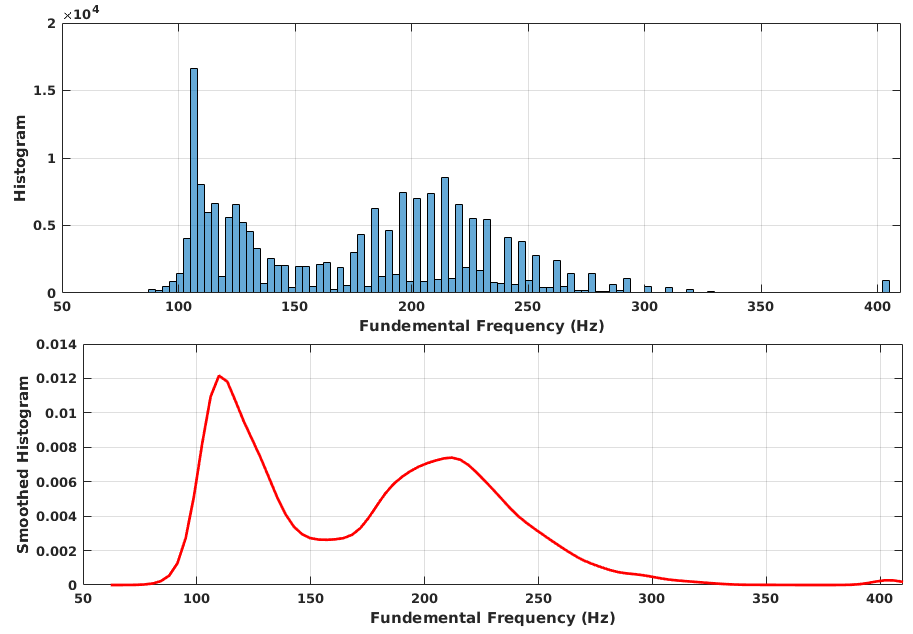}
\caption{Distribution of the fundamental frequency estimates from a PLTL session that consists of eight audio streams of 80 minutes duration. We dropped the non-speech frames (that was assigned a fundamental frequency of zero (0) Hz.}
\label{fig_pitch_hist}
\end{figure}
This section describes the robust pitch extraction using deep neural network trained on stacked spectral features (Pitch Estimation Filter with Amplitude Compression)~\citet{gonzalez2014pefac}. The pitch estimates were later used for detecting curiosity (in terms of question-inflection) and emphasized speech. 

We tested various pitch estimation algorithms such as modified autocorrelation method~\citet{de2002yin}, Sawtooth Waveform Inspired Pitch Estimator~\citet{camacho2008sawtooth}, Subband Autocorrelation Classification~\citet{lee2012noise} and deep neural network (DNN)~\citet{han2014neural}. The state-of-the-art pitch tracking method use a deep neural network (DNN) trained on spectral features~\citet{han2014neural} for predicting the pitch states. DNN-based pitch tracker was the best among four alternatives we tested. The parameters of system used for pitch extraction is given in Table~\ref{table_params_pitch}. 

\begin{figure}[!t]
\centering
\includegraphics[width=240pt]{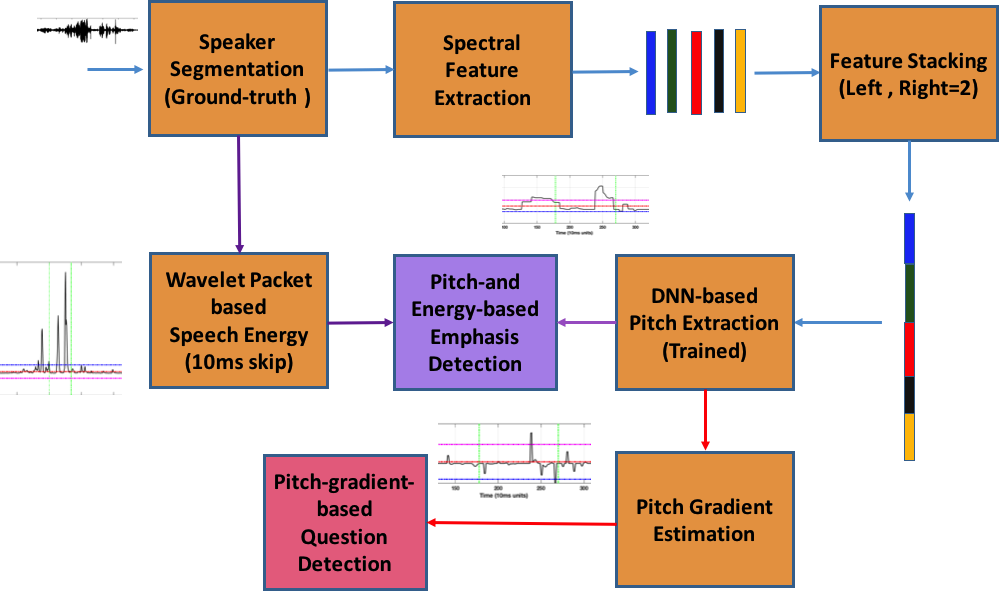}
\caption{Block diagram of the proposed method for detecting curiosity and emphasis in PLTL sessions. Frame-wise pitch was extracted using a deep neural network trained on stacked spectral features (Pitch Estimation Filter with Amplitude
Compression)~\citet{han2014neural}. The pitch information along with speech energy was used for detecting the emphasized regions. The pitch gradient was used for detecting the question-inflection (a measure of curiosity).}
\label{fig_pitch_blk}
% \vspace{-5mm}
\end{figure}

We would briefly cover the DNN-based pitch estimator adopted from~\citet{han2014neural}. The stacked spectral features (Pitch Estimation Filter with Amplitude Compression)~\citet{gonzalez2014pefac} were used to train three-hidden-layer DNN to learn the pitch states. Viterbi decoding was used to connect the probabilistic pitch states, thus fetching the pitch contours. DNN pitch tracker was robust to high amount of noise and worked well for PLTL data. Authors compared the accuracy of DNN pitch tracker with other methods in~\citet{han2014neural}. Spectral features used for training DNN (See Figure~\ref{fig_pitch_blk}) were developed in~\citet{gonzalez2014pefac}. The log-frequency power spectra was normalized to capture long-term information and further filtered to suppress the noise and enhance the harmonic structure in speech frames~\citet{gonzalez2014pefac}. 

Pitch Estimation Filter with Amplitude
Compression features were earlier used for pitch tracking in noise by peak-picking~\citet{gonzalez2014pefac}. These features were stacked using two past and two future frames as shown in Figure~\ref{fig_pitch_blk} (see Table~\ref{table_params_pitch}). The reverberation and noise in CRSS-PLTL data posed challenge for pitch extraction that necessitated use of DNN-based pitch tracker. We smoothed the DNN extracted pitch using Savitzky-Golay filter~\citet{schafer2011savitzky} with third order and 11 frames. The smoothing helped in further correction of pitch values for PLTL data. Figure~\ref{fig_pitch_hist} show the distribution of pitch estimates obtained using DNN-based system. It was obtained on a 80 minute audio from a PLTL session. DNN could accurately estimate the pitch eliminating the pitch doubling that was common in unsupervised methods for pitch estimation. The non-speech frames (corresponding to fundamental frequency of 0 Hz) were dropped for plotting this distribution.
\section{Robust Speaker Diarization}
\begin{figure*}[!t]
\centering
\includegraphics[width=450pt]{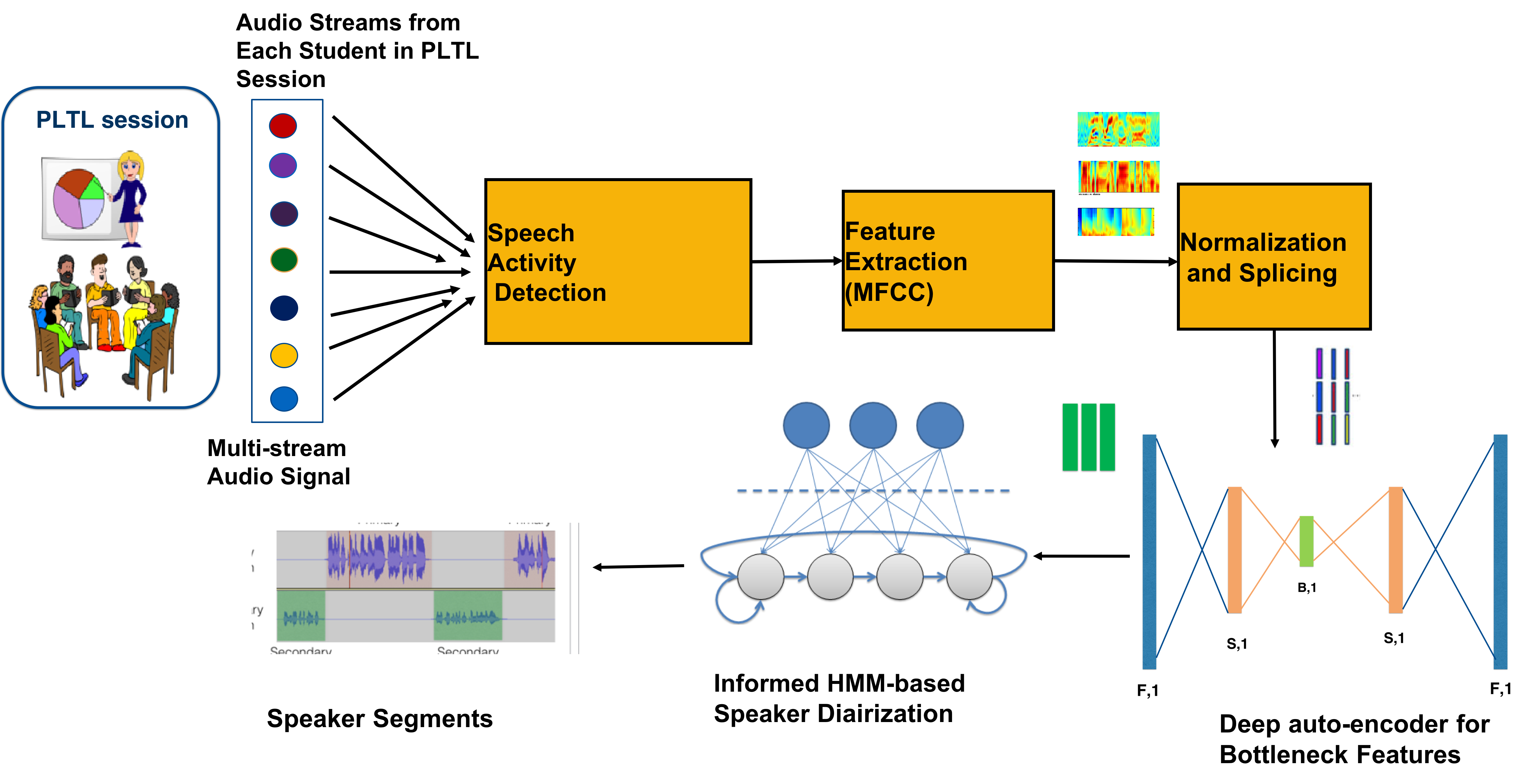}
%\vspace{-2mm}
\caption{Proposed diarization system. It has two main components: (1) stacked autoencoder based bottleneck features that incorporated splicing with context of five past and future frames and takes acoustic features from all streams of PLTL data; (2) Informed HMM-based diarization system that incorporated the number of students (same as number of audio channels) and minimum duration of conversational-turns as side information.}
\label{fig_diar}
\end{figure*}
\subsection{Baseline LIUM Diarization System}
\label{sec:baseline}
We used the LIUM speaker diarization system as the baseline~\citet{meignier2010lium}. The standard LIUM system has poor performance as shown in Table~\ref{table_results_der}. Sampling rate of 8 kHz was used for all experiments reported in this paper. The speech data was divided into 25ms frames with 10ms skip-rate. LIUM system uses ILP clustering on i-vectors. LIUM performed best on broadcast news recordings where turns were significantly longer than $shorter-than-1-second$ turns in PLTL. The reason LIUM could not accurately segments the speakers in PLTL data is due several factors such as (1) short conversational-turns; (2) reverberation; (3) non-speech such as noise, laughter~\emph{etc.}; (4) overlapped speech.

We used diarization error rate (DER) as the figure of merit for diarization systems. DER, as defined by the NIST Rich Transcription Evaluation~\citet{nistder}, could be computed as, 
% \vspace{-1mm}
\begin{equation}
DER = \frac{L_{f} + L_{m} + L_{e} }{L_{t}}
\label{eqn_der}
\end{equation}
% \vspace{-1mm}
where $L_{f}$ was the total number of non-speech segments detected as speech(false alarm), $L_{m}$ is the total number of the speech segments detected as non-speech (miss), $L_{e}$ is the total number of speech frames that were clustered as incorrect speakers (error in clustering), and $L_{t}$ is the total number of ground-truth speech frames. 
\subsection{Stacked Autoencoder-based Bottleneck Features for Diarization}
\label{sec:dae}
The proposed scheme was depicted in Figure~\ref{fig_diar}. Deep neural network (DNN) could be used for dimension reduction for high dimensional feature vectors~\citet{hinton2006reducing}. Autoencoders were found useful in dimension reduction task~\citet{wang2016auto}. This network was trained in a way that allowed it to learn low-dimensional hidden representation of the data such that taking noisy input, it could reconstruct the input. 

Input feature vectors were corrupted with additive random noise. We used 13 dimensional Mel-Frequency Cepstral Coefficients (MFCC). Each feature dimension was mean and variance normalized. We performed splicing of normalized feature vectors by taking five past and future frames. The stacked autoencoder was used for extracting the bottleneck features (bottleneck) from spliced and normalized MFCC features. Several autoencoders were stacked to form a deep network with five layers. Stacked autoencoder was trained using spliced features. Stacked autoencoders were first trained in layer-wise fashion that is a standard way of pre-training. After pre-training, stacked autoencoder was fine-tuned so that it could reconstruct the input features. The input to the stacked autoencoder was corrupted before feeding into it. The reconstruction-loss was minimization criterion for training this network~\citet{vincent2008extracting}. 

We used PDNN toolkit~\citet{miaopdnn} with corruption parameter 0.2, learning rate, and momentum factor parameters of 0.01 and 0.05, respectively for realizing the stacked autoencoder. The parameters of the stacked autoencoder used for bottleneck feature extraction was given in Table~\ref{table_params_diar}. The feature vectors (13-MFCC) were first mean and variance normalized. Let $\mathbf{m}$ was the feature vector, $\mathbf{\mu_{m}}$ and $\mathbf{\sigma_{m}}$ were the mean and standard deviation vectors, respectively. The normalized feature vector, $\mathbf{\bar{m}}$, is given by $\mathbf{\bar{m}}= \frac{\mathbf{m} - \mathbf{\mu{m}}}{\mathbf{\sigma_{m}}}$. 

Since all the channel were delayed and scaled versions of the same speech signal at a given frame, using all channels for diarization was important. Time-spliced feature vectors from each channel were concatenated to form a supervector that consisted of feature vectors corresponding to all PLTL channels. The room where PLTL data was collected has dimensions of 7X10 meters. Thus, the maximum distance between a LENA device and any students (other than the wearer) can be assumed to be within ten meters. Taking the speed of sound in air to be 343 meters per second (m/s), we have the maximum time delay, to be of the order 30ms. This calculation did not accounted for reverberation. We used 25ms windows with 10ms skip-rate for our experiments as given in Table~\ref{table_params_diar}. We concatenated the features from all streams. The normalized feature super-vectors were spliced by taking five past and future frames. The concatenation was done to incorporate time and intensity differences between various channels of multi-stream PLTL data. The splicing incorporates the long-term context leading to a better quantification of reverberant and noisy speech frames. For a PLTL group with seven streams, the final dimension of spliced features was 11*7*13-MFCC, i.e., 1001.

\subsection{Informed-HMM based Diarization System}
\label{sec:diar}
In this section, we discuss informed-Hidden Markov Model (HMM) for joint speaker segmentation and clustering. HMM system incorporate the bottleneck features from stacked autoencoder system~\citet{gehring2013extracting} along with two dimensions of side information,~\emph{i.e.}, (1) number of speakers; and (2) minimum duration of conversational-turns. Hence, we called the system as informed HMM system. The iterative diarization procedure had three steps: (i) initial segmentation, (ii) merging, and (iii) re-estimation. 

The diarization for PLTL sessions was different with respect to information available such as speaker-count and turn-statistics. The rapid short-turns, overlapped-speech and significant noise and reverberation made the task challenging. Most of the studied diarization system did not address such challenges~\citet{dubey2016interspeech,anguera2012speaker}. PLTL sessions had frequent short-segments of size 0.2s to 1s and few segments of size 1-3s. HMMs had been used in previous studies for various audio segmentation tasks in varied forms~\citet{fredouille2006technical,madikeri2015kl,kotti2008speaker,ajmera2002improved,huang2006advances}. However, using side information, application to PLTL and using stacked autoencoder-based bottleneck features were novel contributions with respect to speaker diarization.

Initially, we performed over-segmentation by dividing speech into $OS$ segments where $OS$ was four to six times the expected number of speakers. A HMM with $OS$ states was assumed for initial segments. Each HMM state had an output probability density function that was modeled by $M$ component Gaussian Mixture Model (GMM). Each state of HMM was allowed to have $T$ sub-states to incorporate the minimum duration constraint. All sub-states of a given HMM state (hypothesized speaker cluster) share the GMM corresponding to their state. The HMM system was trained using~\textit{Expectation-Maximization} (EM) algorithm. One step aimed to segment the data such that their likelihoods given corresponding GMM parameters were maximized. In next step, the GMM parameters were re-estimated based on new segmentation. Once HMM was trained, we obtained the Viterbi path for each frame. Following it, we used the Viterbi path for checking the binary merging hypothesis based on modified $G^{3}$ algorithm~\citet{dubey2016interspeech}. After the merge iteration finished, a new HMM with less number of states was trained. The whole process was repeated again till the number of HMM states equaled the number of speakers. 

We performed merging based on $G^{3}$ algorithm that was a variant of BIC and eliminated the need of the penalty term. The unsupervised $G^{3}$ algorithm~\citet{dubey2016interspeech} was used for deciding the binary hypothesis of merging two segments. This trick was first developed to improve the speaker change detection as compared to BIC~\citet{ajmera2004robust}. In this paper, we used the same techniques for a different binary hypothesis to decide merging of two over-segmented segments or equivalently two HMM states. There are some modifications to $G^{3}$ algorithm applied for merging most-similar segments (HHM states) at each iteration of the informed HMM-based diarization system. First, the minimum duration of staying in a HMM state was much lower, 0.2s to 0.5s owing to the rapid short conversational-turns. The initial segments were modeled with a Gaussian Mixture Model (GMM) with only $M_{s}$ components. After merging two initial segments modeled with $M_{s}$ components, the merged segment was modeled with $2M_{s}$ components. Thus, the number of parameters in GMM model for merged segment is same as the sum of number of parameters in child segments. Consequently, the number of parameters remains the same at each merging step, and hence the penalty term in BIC criterion (See Equation~\ref{eqn_bic}) is eliminated.

Let $\mathbf{X_m}= [\mathbf{X_{1}}, \mathbf{X_{2}}] $ be the feature matrix corresponding to the merged HMM states. Merging two segments, $\mathbf{X_{1}}$ and $\mathbf{X_{2}}$ into $\mathbf{X_m}$ can be formulated as the following binary hypothesis: $\mathcal{H}_{0}$ \emph{vs.} $\mathcal{H}_{m}$, where $\mathcal{H}_{m}$ denotes merging, and $\mathcal{H}_{0}$ denotes no merging. To facilitate the test, we build models for both hypotheses. GMMs were used to model $\mathbf{X_{1}}$ , $\mathbf{X_{2}}$ and merged segment $\mathbf{X_{m}}$. Let $\psi_{X_{m}}$ be the parameter vector of the GMM with $M_{s} = M_{1} + M_{2} $ component estimated for the merged segment, $\mathbf{X_{m}}$. Let, $\psi_{X_{1}}$ and $\psi_{X_{2}}$ be the parameter vector of the GMMs with $M_{1}$ and $M_{2}$ components, estimated for the child segments, $\mathbf{X_{1}}$ and $\mathbf{X_{2}}$, respectively. Under the assumption of independence and identical distribution of feature vectors in segments $\mathbf{X_{1}}$ and $\mathbf{X_{2}} $, we can represent the log likelihood $\mathcal{L}_{\mathcal{H}_{0}}$ and $\mathcal{L}_{\mathcal{H}_{m}}$ for hypotheses $\mathcal{H}_{0}$ and $\mathcal{H}_{m}$, respectively as
%%%
\begin{equation}
\mathcal{L}_{\mathcal{H}_{m}} = \log(p(\mathbf{X_{1}}|\psi_{X_{m}})) + \log(p(\mathbf{X_{2}}|\psi_{X_{m}})) , 
\end{equation}
\vspace{-1mm}
\begin{equation}
\mathcal{L}_{\mathcal{H}_{0}} =  \log(p(\mathbf{X_{1}}|\psi_{X_{1}})) + \log(p(\mathbf{X_{2}} |\psi_{X_{2}})) ,
\end{equation}
where $p(\mathbf{X_{m}}|\psi_{X_{m}})$ is the likelihood of merged segment, $X_{m}$  given the model, $\psi_{X_{m}}$, and so on. The merging decision is made based on the $D_{merging}$, defined as 
% \vspace{0mm}
\begin{equation}
D_{merging} = \mathcal{L}_{\mathcal{H}_{m}} - \mathcal{L}_{\mathcal{H}_{0}} ,
\label{eqn_merge}
\end{equation}
However, if we used Bayesian Information Criterion (BIC) for making the merging decision, then corresponding to Equation~\ref{eqn_merge}, we have following expression for BIC merging:
\begin{equation}
D_{BIC} = \mathcal{L}_{\mathcal{H}_{m}} - \mathcal{L}_{\mathcal{H}_{0}} - \frac{1}{2}\nu \Delta \log N_{m} ,
\label{eqn_bic}
\end{equation}
where $\nu$ is a constant usually assigned a value of $1.0$ and $N_{m}$ is the number of feature vectors in merged segment, $\mathbf{X_{m}}$. Here, $\Delta$ is the difference in number of parameters in merged model, $\psi_{X_{m}}$ and sum of parameters in child models, $\psi_{X_{1}}$ and $\psi_{X_{2}}$. All segments were evaluated for $D_{merging}$. The segments with $D_{merging} \ge 0$ were merged. Once the merging  done, the new HMM of smaller size was estimated where the GMM for each state was re-estimated using the EM algorithm. The acoustic features belonging to that HMM state (cluster) were used to re-estimate the corresponding GMM.

\section{Behavioral Characteristics}
\label{sec:behave}
\begin{figure}[!t]
\centering
\includegraphics[width=230bp]{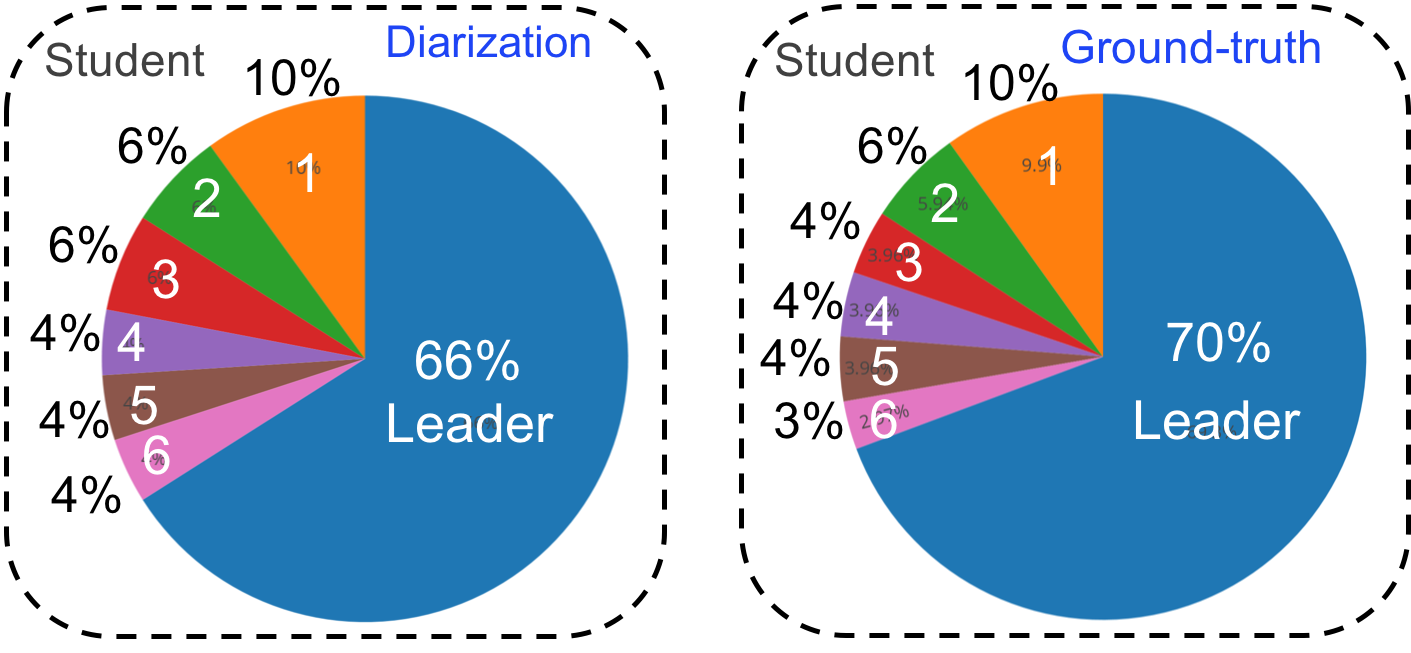}
\caption{Participation analysis of Eval-Set-2 (see Table~\ref{table_eval_sets}) that consisted of 21 minute audio data. It depicts the percentage time for which each individual was speaking. We can see all students occupy comparable fraction of conversation floor while the peer leader occupied the highest fraction.}
\label{fig_participation}
\end{figure}
\subsection{Participation Analysis}
\label{sec:participation}
Diarization output could be used for extracting participation analysis, that refers to the percentage of total time for which each speaker and their team leader occupied the conversation floor. Figure~\ref{fig_participation} shows the participation analysis obtained using 21 minutes of data, Eval-Set-2 (See Table~\ref{table_eval_sets}) from a PLTL session.  The comparison between diarization-based participation analysis and ground-truth clearly shows that even if diarization error rate is non-zero, we can still derive meaningful participation analysis from it. The percentage values were rounded-up for better visualization.
\subsection{Dominance Score}
\label{sec:dominance}
Dominance in human-to-human communication had been studied for several decades~\citet{young2016handbook}. Dominance is a fundamental aspect of interactions in PLTL sessions. The researchers in social psychology have studied dominance in human interactions~\citet{dunbar2005perceptions}. The speaking time of speakers were found to be correlated with perceived dominance of individuals in groups~\citet{mast2002dominance}. Researchers in social signal processing studied dominance models developed from multi-modal data.

Authors measured the dominance in meeting using the speaker diarization techniques~\citet{hung2008estimating}. Authors developed a supervised model for dominance using short-utterances~\citet{basu2001learning}. However, the model was developed and evaluated on a constrained settings that was very different from real-life situations such as PLTL sessions. Authors analyzed the interaction between two individuals who debated for 60 seconds. Such controlled settings and short-duration analyses were not applicable for spontaneous conversations such as those in PLTL sessions. Authors used multi-modal features derived from audio and video streams for analyzing the dominant persons in meetings~\citet{hung2007using}. 

Authors used manual transcriptions of meetings for generating semantic metrics that were later used for training static and dynamic models of dominance~\citet{rienks2006detection}. However, they did not process the audio rather the text was processed to build the supervised models. Such systems could not be deployed for analysis of PLTL groups as they required scripting and training supervised classifiers. Authors  proposed a dominance model for meetings based on supervised learning using multi-modal data (multi-microphone audio and multi-camera video). The audio and visual data were used for training support vector machine classifier. It was used for training the supervised dominance model for meeting conversations~\citet{jayagopi2009modeling}. However, such a system need supervised training on huge amount of labeled multi-modal data and was likely to perform poorly under mismatched conditions. Another limitation was that it could not be used if only audio data were available from PLTL sessions.

We developed an unsupervised feature for measuring dominance~\citet{dubey2016robust}. Dominance score (DS) was assigned to each student by unsupervised acoustic analysis of their speech segments. The proposed DS encapsulates the probability of a given student to be dominant in collaborative problem solving. We considered three features derived from speech corresponding to each speaker. This information was available from informed HMM-based diarization system (see Figure~\ref{fig_diar}). 

The three features are turn-taken-sum ($turns$)~\citet{larrue1993organization}, speaking-time-sum ($spts$), and speaking-energy-sum ($spens$). These features were motivated from social psychology literature where the dominance of a speaker was found to be correlated with taking more turns in a conversation, speaking for longer duration~\citet{mast2002dominance}, and with higher energy~\citet{dunbar2005perceptions}. 

These features were correlated among themselves. For example, a person who was taking many turns was likely to speak for longer duration than others. Also, adding the speaker energy for a longer duration would result in higher $spens$. The turn-taken-sum ($turns$) was the number of turns taken by the speaker in a given session. A conversation turn was decided by a speech segment from the speaker cascaded between that from other speakers and/or speech pauses (non-speech). The speaking-time-sum ($spts$) was the sum of duration of all time-segments (in seconds) belonging to that speaker. The overlapped speech was not taken into account in this sum. Speaking-energy-sum ($spens$) was sum of energies for that speaker's segments. 

The speech energy was computed using wavelet packet decomposition~\citet{wickerhauser1991lectures} as discussed in Section~\ref{sec:wpd}. The PLTL data had huge reverberation and noise, that necessitated development of better metric for computing speaking energy. We used the Symlets6(sym6) wavelet with six levels of decomposition for computing the speech energy. The coefficients corresponding to frequency range [50, 2000] Hz were summed to get the energy. 

After extracting these three features, $turns$, $spts$ and $spens$, we normalized each feature dimension. Let $\mathbf{f}$ be the three dimensional feature-vector, $\mathbf{\mu}$ and $\mathbf{\sigma}$ being the mean vector and standard deviation vector. The normalized feature vector, $\mathbf{\bar{f}}$, is given by $\mathbf{\bar{f}}= \frac{\mathbf{f} - \mathbf{\mu}}{\mathbf{\sigma}}$. Here, the division is point-wise, the mean and variance were calculated over the entire PLTL session (approximately 70-80 minute audio). 

We projected these normalized features onto eigen space corresponding to the highest eigen value of the feature space. This was realized by principal component analysis (PCA) that combined the three features into a single feature, named $comb$ feature (short form for combined feature). Let us denote the $comb$ feature by $p$. We computed the $comb$ feature for each speaker in each segment of the PLTL session. PCA was performed for the whole PLTL session. In this paper, we divided the entire PLTL session into five-minute segments. A dominance score was estimated for each speaker during five-minute segments.

Lets say, $p_{i}$ was the $comb$ feature corresponding to $i-th$ speaker. For CRSS-PLTL corpus we have six to nine speakers in sessions including team leader. We defined $comb$ feature-vector as, $\mathbf{p} = [p_{1}, p_{2},..,p_{N}]$, where $N$ was the number of speakers. The dominance score (DS) for each speaker was estimated by processing the dominance feature-vector, $\mathbf{p}$, through a soft-max function that convert these numbers into probability scores. Thus, we had
\begin{equation}
DS_{i}= \frac{e^{p_{i}}}{\sum_{j=1}^{N}e^{p_{j}}}, 
\label{eqnd2}
\end{equation}
for $i=1,2,..,N$; where $DS_{i}$ was the dominance score (DS) of the $i-th$ speaker.

Once we have the dominance score, finding the most and least dominant speaker was trivial. The one with highest score was the most dominant person while the one with lowest was least dominant. In PLTL sessions, the dominance score of each students is an important metric with respect to inter-session variability for all sessions of that team. From previously studied supervised dominance models that predicted only the most dominant speaker, such a comparison would not be possible~\citet{jayagopi2009modeling,hung2007using,huang2006advances}. Dominance analysis could help in understanding the role of each team member in a PLTL session with respect to learning of their own and others. It could help in choosing suitable candidates for a PLTL session  so as to maximize the learning outcome for each one of them.
\subsection{Curiosity: Question-Inflection Detection}
\label{sec:curiosity}
\begin{figure}[!t]
\centering
\includegraphics[width=240pt]{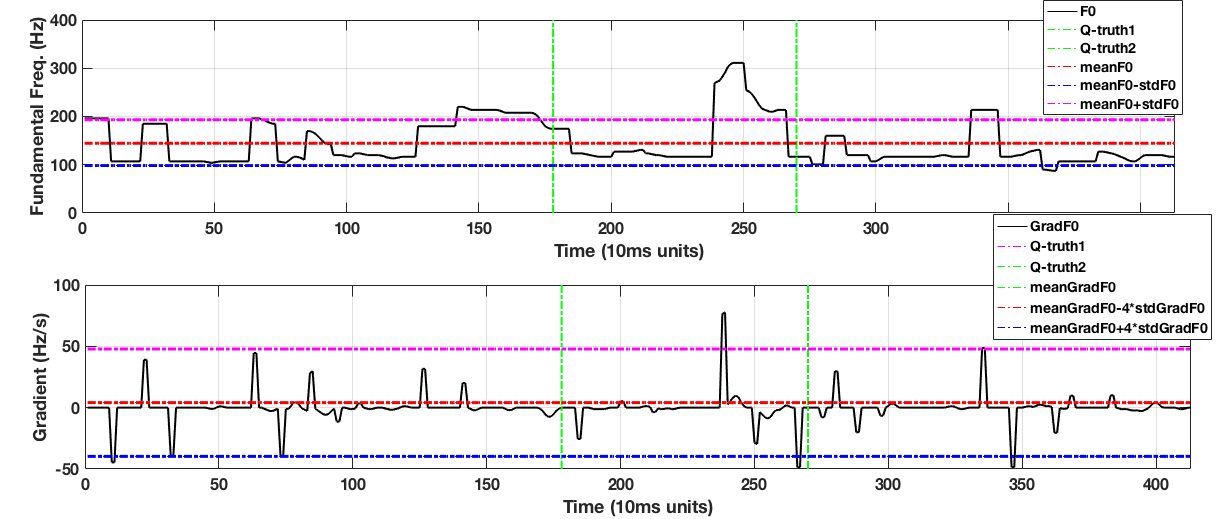}
\caption{Detection of question-inflection using gradient of pitch contour. The top sub-figure shows the pitch contour along with start-time (Q-truth1) and end-time (Q-truth2) of the question-inflection, and mean-Pitch $\pm$ std-Pitch lines. The bottom sub-figure shows the gradient of pitch contour along with mean, and meanGradPitch $\pm$ 4*stdGradPitch lines. We could see that question-inflection was accompanied by low-to-very high pitch inflation leading to a local maxima at the end of the question (see top sub-figure). We detect the question-inflection by a statistical rule as shown in bottom sub-figure. The frames that belong to GradPitch $\ge$ meanGradPitch $\pm$ 4*stdGradPitch corresponds to a question-inflection.}
\label{fig_Qdetect}
\end{figure}
 \begin{figure}[!t]
 \centering
 \includegraphics[width=240pt]{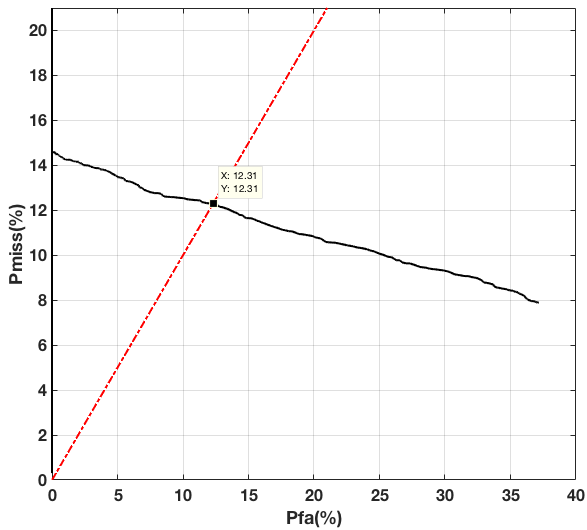}
 \caption{Detection error trade-off (DET) curve for 30 minutes of audio data for question-inflection detection. The pitch contour from complete signal was mean and variance normalized over non-overlapping 2-second segments. The equal error rate (EER) comes out to be 12.31\%. The threshold for detection of question was varied to determine various points (each point corresponds to a miss rate and false alarm rate) shown in this curve.}
 \label{fig_eer}
 \end{figure}
Curiosity refers to a desire for gaining new information or skill~\citet{renner2006curiosity}. Curiosity was defined in the study as "aurally identifiable trait of the internal desire" of PLTL participants to acquire new information or skills. The curiosity is an important trait in learning~\citet{renner2006curiosity}.  A pitch transform was used for synthesis of interrogative sentence in~\citet{nagy2016improving}. 

Eval-Set-5 (30 minute audio data) was used for evaluating the algorithm for question-inflection detection. The audio data was annotated for start-time and end-time of each question. The annotation was done over five minute super-segments. The time-stamps for each question-inflection were located. Gradient of the pitch contour for each speaker segment was computed to find the local maximum. Question inflection was detected when the pitch gradient goes above the value of $meanGradPitch + 4*stdGradPitch$ (the mean and std are computed using gradient contour over that segment).

We annotated the start-time and end-time of the segment when question was asked. The mid-point of ground-truth question-boundary was used for computing correlation and root mean squared error with algorithm detected question-inflection point. Figure~\ref{fig_Qdetect} shows the pitch variations on a question onset and its neighborhood. It also shows the gradient contour and detection of question-inflection. 

We designed another experiment to study the pitch-based question-inflection detection. We took the evaluation audio data and estimated the pitch contour for complete signal regardless of speaker-change boundaries. We performed the mean and variance normalization of pitch contour over each two-second non-overlapping segments. Normalization compensated the long-term effects making the pitch contour robust to acoustic variability. Normalized pitch was used for detecting the question-inflection by choosing a threshold. We varied the threshold from minimum to maximum value (of pitch contour) in small steps. For each threshold values, we get miss probability, $Pmiss$ and probability  of false alarm , $Pfa$ (in \%) with respect to detection of question-inflection. 

For EER calculation (DET curve), all frames belonging to the time-interval during which a question was asked, were taken as question-inflection points. This is different from root mean squared error and correlation computation where the mid-point of ground-truth question-boundary was compared with point of question-inflection detection. Figure~\ref{fig_eer} shows the detection error trade-off (DET) curve for Eval-Set-5 data. The equal error rate (EER) was 12.31\%. Here, $Pmiss$ refers to the frames where we had the questions asked but the system failed to detect it (miss). $Pfa$ refers to the frames where question-inflection was falsely detected (false alarm). In this paper, we used only single channel data for annotation and evaluation for pitch-based question-inflection detection for simplicity in evaluation. 

\subsection{Emphasis Detection}
\label{sec:emphasis}
\begin{figure}[!t]
\centering
\includegraphics[width=230pt]{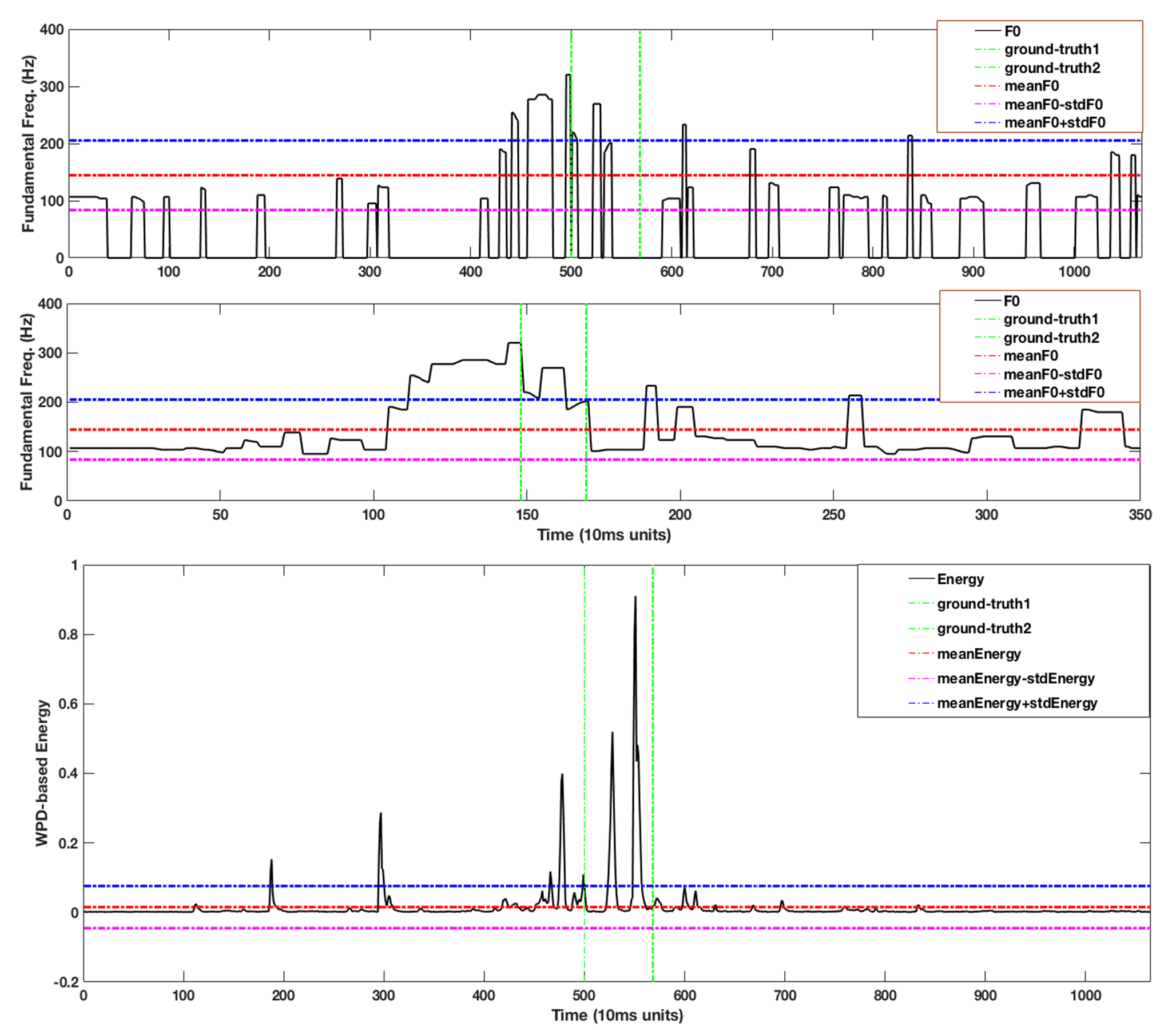}
\caption{Showing pitch-and energy-based emphasis detection. Top sub-figure showed the pitch contour for a speaker segment with emphasized region. The middle sub-figure showed only speech frames (with non-zero pitch). Bottom sub-figure showed the frame-level speech energy obtained using wavelet packet decomposition. When the pitch was higher than $meanPitch$ + $std-Pitch$ and energy was higher that $meanEnergy$ + $stdEnergy$, the emphasized region was detected.}
\label{fig_pitchEmp}
\end{figure}
\begin{figure}[!t]
\centering
\includegraphics[width=230pt]{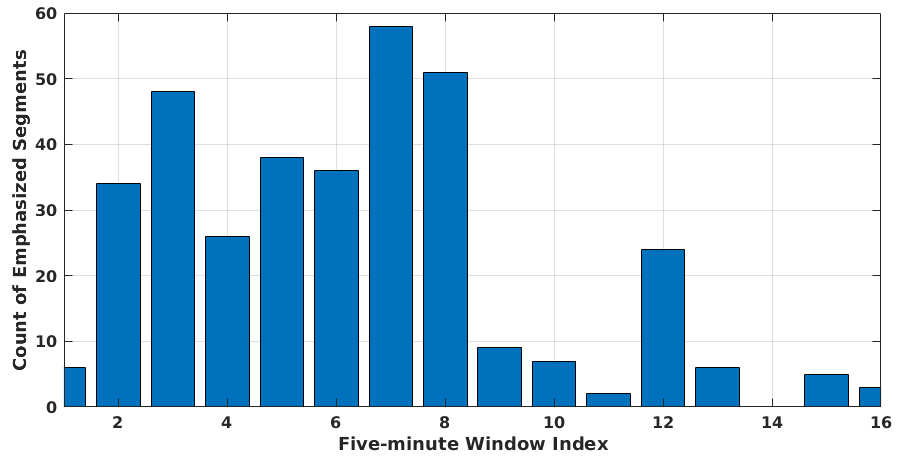}
\caption{We chose a PLTL session with eight students that was organized for 80 minutes. We divide the session into five-minute segments. This bar graph shows the number of emphasized speech regions in each of these five-minute segments. We could be observed that the highest number of emphasized segments occurred around the middle of the session. The last segments were more about logistics and general questions \& answers that did not involve emphasized regions.}
\label{fig_emp_count}
\end{figure}

Detection of emphasized speech could help in discovering the "hot-spots" in PLTL sessions wherein important discussions might have happened. Such segments could help education researchers in understanding and designing the best practices. Student's excitement could be captured by detecting such segments. Emphasized speech regions were important with respect to semantic analysis. Such segments could be further processed with natural language processing (NLP) tools.  We have the option of using NLP tools on complete session, however using NLP only on few emphasized segment could reduce computations by eliminating segments that were relatively less important. We used the pitch contour and speech energy for detecting the emphasized speech. These regions identify the important regions in audio data. 

The emphasis detection from audio had been studied previously~\citet{chen1992use,arons1994pitch,arons1992techniques,arons1997speechskimmer}. Detecting the emphasized regions helped in quick summarization of spoken documents~\citet{arons1997speechskimmer}. Such summaries collected the salient features of the recordings and were useful for analysis of technical discussions and daily-life conversations. A HMM-based model trained on huge amount of data was used for emphasis detection in~\citet{chen1992use}. Pitch changes were leveraged for detection of emphasized regions in meetings~\citet{kennedy2003pitch}. 

However, the past works~\citet{chen1992use,arons1994pitch,arons1992techniques,arons1997speechskimmer} had not been tested over long-duration spontaneous speech with several speakers (such as six to eight participants in PLTL session). CRSS-PLTL data had short conversational-turns at several instances in addition to noise and reverberation, thus making the task challenging. Since we estimated the pitch contour and do the analysis for each speaker segment, the pitch range of each speaker is automatically taken into account. As the pitch could change abruptly due to speaker changes (for example, from a male to female speaker), it was important to have accurate speaker segments. The proposed algorithm adapted to the pitch and energy range of a speaker (by operating over non-overlapping two-second windows), and then automatically selected the regions of increased pitch-and energy-activity as a measure of emphasis. Increase pitch and speech energy are markers of an emphasized region while pitch information was found to be more important~\citet{chen1992use}. 

We proposed detection of emphasized speech using inflated speech energy and increased pitch. The wavelet packet decomposition was used for robust estimation of speech energy as explained in Section~\ref{sec:wpd}. The correlation and root mean squared error between ground-truth (mid-point) and estimated point of emphasis detection were used as figure of merit for this method. Figure~\ref{fig_emp_count} showed the distribution of emphasized regions in each five-minute segments of a PLTL session (approximately 80 minutes).  We could see the highest number of emphasized speech segment lies in mid of the session. It showed that the "hot-spots" in PLTL sessions were more often during the mid-time. 

Figure~\ref{fig_pitchEmp} shows detection of emphasized segments using pitch and energy. Emphasis was detected based on two conditions: 1) energy higher than $meanEnergy + stdEnergy$, and 2) pitch higher than $meanPitch + std-Pitch$. Simultaneous satisfaction of these conditions detected emphasized speech regions. We had the start-time and end-time boundaries for emphasized regions from manual annotation as described in Section~\ref{sec:annotation}. We took the mid-point of ground-truth emphasis-boundary and estimated its correlation with algorithm computed point of emphasis detection. Also, we calculated the root mean squared error (in units of second), between these two quantities, i.e., ground-truth and estimated detection point. Table~\ref{table_results_pitch} shows the evaluation results. 

\begin{figure*}[!t]
\centering
\includegraphics[width=450pt]{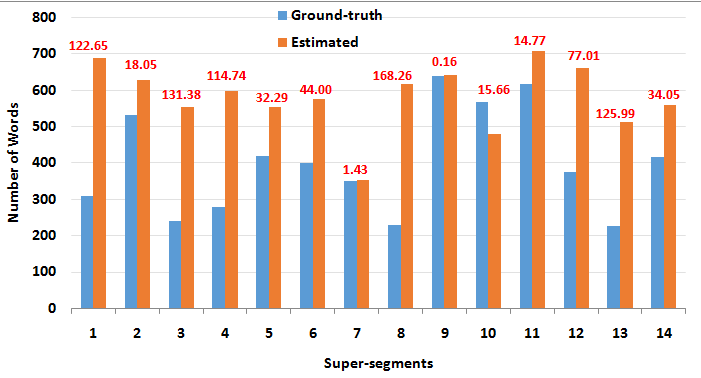}
\caption{The word count ground-truth and estimated using~\citealt{ziaei2016effective} for Eval-Set-6 (see Table~\ref{table_eval_sets}) that consisted of 70 minutes of audio data. We could see that performance varies from very good to very poor. It depicted the changing acoustic scenarios that affected the quality of the speech signal. The red number above the bars showed the percentage error rate with respect to ground-truth word count. The low errors occur when speaker wore the LENA device and high error occurred due to voice of a distant speaker. The reverberation levels were different for each unique position of the speaker. Very low error in seventh and ninth segment showed that method worked well when speech quality was good and very high errors in first, third and thirteenth segment shows that method proposed in~\citet{ziaei2016effective} got worse when speaker changes were rapid and/or some of the speakers were far from the LENA device.}
\label{fig_wc}
\end{figure*}
\subsection{Engagement: Speech Rate}
The speech rate was considered an important aspect of vocal communication~\citet{cummins2009rhythm}. Speech rate was useful for quantifying the engagement behavior. Increased speech rate showed more engagement. Authors used prosodic cues for studying engagement behaviors in children~\citet{gupta2016analysis}. Several interaction scenarios between a child and psychologist were used for validating the developed algorithms. Engagement was predicted using vocal and prosodic cues. Authors concluded that the engagement information was not only reflected in global cues but also in short-term local cues. Three levels of engagement were used for experimental validation. Fusing global and local cues gave the best results. Even though the experiments were validated in constrained settings, it showed that certain prosodic patterns captured the engagement in dyadic interactions~\citet{gupta2016analysis}. 

Several algorithms were developed for estimating the speech rate~\citet{morgan1998combining,jiao2015estimating,wang2007robust,ziaei2016effective}. We benchmarked the method developed in~\citet{ziaei2016effective} on Eval-Set-6 (see Table~\ref{table_eval_sets}) derived from the CRSS-PLTL corpus. It consisted of 70 minute audio from a PLTL session. 

Figure~\ref{fig_wc} shows the evaluation of word count algorithm~\citet{ziaei2016effective} on Eval-Set-6. We divided the PLTL session into five-minute segments and performed the word count estimation using method proposed in~\citet{ziaei2016effective}. The red numbers above the bars showed the percentage error rate with respect to ground-truth word count. We could see the performance varying from very low to high error rate. The low errors occurred when speaker wore the LENA device and high error was possibly due to the speech of a distant speaker and rapid short-turns from several speakers (six to eight student were in a PLTL session). The reverberation levels were different for each unique position of speakers. Very low error in seventh and ninth segments showed that method worked well when speech quality was good and very high errors in first, third and thirteenth segment shows that method in~\citet{ziaei2016effective} got worse when speaker changes were rapid and some of the speaker were far from the LENA device. It showed the necessity to investigate reverberation-and noise-robust methods for speech rate estimation that could work accurately for naturalistic audio streams.
\begin{table}[!t]
\centering
\caption{The parameters set for proposed diarization system that consisted of three main parts: (1) acoustic feature extraction, (2) stacked autoencoder (autoencoder)-based bottleneck features, and (3) informed HMM-based diarization system.}
%\vspace{-2mm}
%\begin{tabular}{*{3}{|c|}}
%\begin{tabular}{*{2}{|c|c|}}
\begin{tabular}{|c|c|}
\hline
\textbf{Parameter} & \textbf{Value}\\
\hline
Stacked autoencoder input layer dim.&1001\\ \hline
Stacked autoencoder second layer dim.&91\\ \hline
Stacked autoencoder bottleneck layer dim.&21\\ \hline
Number of hidden layers~\ref{fig_diar}&3\\ \hline
First layer activation&tanh\\ \hline  
Hidden layer activation&sigmoid\\ \hline  
Initial states in HMM&12-18\\ \hline
Number of GMM components&2-5\\ \hline
Minimum duration of HMM states&0.2s-1s\\ \hline
Splicing context (past)&5 frames\\
\hline
Splicing context (future)&5 frames\\
\hline
Features&MFCC\\ \hline
Window length&25ms\\\hline
Skip-rate&10ms\\ \hline
Sampling rate &8000Hz\\ \hline
\end{tabular}
\label{table_params_diar}
\end{table}
%% sec6
%
%%
\begin{table*}[!t]
\centering
\caption{Comparison of Diarization Error Rate (DER) for various parameters of the stacked autoencoder-based bottleneck features and Informed HMM-based diarization system. $I_{K}$ is initial number of clusters (hypothesized number of speakers) and $I_{G}$ is the number of Gaussian components in initial model for over-segmented clusters.}
\label{table_results_der}
\centerline{
\begin{tabular}{*{3}{|c|c|c}}
\hline
\textbf{SAD}&$feat_dim$&$t_{min}(s)$&$I_{K}$& $I_{G}$&DER(\%)\\\hline
LIUM~\citet{meignier2010lium}~\ref{sec:baseline} & & & & &35.80 \\ \hline
NO SAD&13-MFCC (* 7= 91 from seven streams)&0.5&12&2&41.71\\ \hline
NO SAD&13-MFCC (* 7= 91 from seven streams)&1&12&2&33.23\\ \hline
NO SAD&19-autoencoder&0.5&12&2&16.64\\ \hline
NO SAD&19-autoencoder&1&12&2&\textbf{15.83}\\ \hline
Oracle&13-MFCC (* 7= 91 from seven streams)&1&12&2&19.98\\\hline
Oracle&13-MFCC (* 7= 91 from seven streams)&0.5&12&2&18.95\\\hline
Oracle&19-autoencoder&1&12&2&\textbf{8.05}\\ \hline
Oracle&19-autoencoder&0.5&12&2&8.87\\ \hline
\end{tabular}
}
\end{table*}
\section{Results \& Discussions}
\label{sec:results}
This section discussed the results obtained on various evaluation sets derived from the CRSS-PLTL corpus. Output of the diarization system was used for getting the participation dynamics as explained in Section~\ref{sec:participation}. Figure~\ref{fig_participation} shows the participation analysis of Eval-Set-2 (see Table~\ref{table_eval_sets}) that consisted of 21 minutes audio data. It was observed that most of the student could speak for comparable time-duration. The team leader occupied the conversation floor for most of the time as peer leader had to facilitate the collaborative problem solving.

\begin{table*}[!t]
\centering
\caption{Speech activity detection was evaluated on Eval-Set-7 data (See Table~\ref{table_eval_sets}) explained in Section~\ref{sec:data-analysis}. Figure~\ref{fig_turn_hist} shows the duration-distribution of speech, non-speech and overlapped-speech segments. Non-speech often contained several noise-sources such as mumbling of far speakers, writing-on-white-board noise in addition to noise from fan and other background sources. We used DNN-based pitch extractor (see Section~\ref{sec:pitch}) for SAD. The frames that were assigned zero(0) pitch were declared non-speech. TO-combo-SAD~\citet{sadjadi2013unsupervised, ziaei2014speech} used voicing measures for detection of speech activity. Later, we fused the output of both SAD systems. If both system's output were not same, we consider those frames as non-speech. Consequently, false alarms were reduced. The non-speech had multiple simultaneous noise-sources that resulted in high false alarm for individual SAD systems. We compared our system with a state-of-the-art SAD system that is a supervised neural network trained over multiple features stream. This system was developed at USC for DARPA RATS data~\citet{van2013robust}. It is interesting to note that comboSAD~\citet{sadjadi2013unsupervised} was an unsupervised SAD algorithm developed for DARPA RATS.}
%\vspace{-2mm}
%\begin{tabular}{*{3}{|c|}}
%\begin{tabular}{*{2}{|c|c|}}
\begin{tabular}{|c|c|c|c|}
\hline
\textbf{System} &  \textbf{System} & \textbf{Pmiss$\%$} & \textbf{Pfa$\%$} \\\hline
(A)& DNN-based pitch &16.54&28.45\\ \hline
(B)& TO-combo-SAD~\citet{ziaei2014speech}& 15.68 &42.64\\ \hline
(C)& Fusion of (A) + (B) & 16.00 &16.64\\ \hline
(D)& USC-supervised-NN~\citet{van2013robust} & 14.68 & 31.70 \\ \hline
\end{tabular}
\label{table_sad}
\end{table*}
\begin{table*}[!t]
\centering
\caption{Showing results for emphasis and question-inflection detection. We used the correlation between ground-truth mid-point and point of emphasized speech-region and question-inflection detection. The evaluation used the oracle speaker segments (except the EER calculation) for question-inflection detection.}
\label{table_results_pitch}
%\vspace{2mm}
\centerline{
%\begin{tabular}{*{3}{|c|}}
\begin{tabular}{|c|c|c|c|}
\hline
\textbf{Quantity}&\textbf{Correlation}&\textbf{root mean squared error(s)}&EER (\%)\\ \hline
Question Inflection&0.84&0.51s&12.31\\ \hline
Emphasis&0.78&0.42s & -- \\ \hline
\end{tabular}
}
\end{table*}
Table~\ref{table_sad} shows the evaluation results on Eval-Set-7 data using methods described in Section~\ref{sec:sad}. Figure~\ref{fig_turn_hist} showed the duration-distribution of speech, non-speech and overlapped-speech segments. Non-speech often contained several noise sources such as mumbling of far speaker, writing-on-white-board noise in addition to noise from fan and other background sources. We used DNN-based pitch extractor (see Section~\ref{sec:pitch}) for SAD. The frames that were assigned zero(0) pitch were declared non-speech. TO-combo-SAD~\citet{ziaei2014speech} was second SAD system used. Later, we fused the output of both systems. For frames where both system's output were not same, we consider those frames as non-speech. Consequently, false alarm was greatly reduced. The non-speech in evaluation set had multiple simultaneous sources that resulted in high false alarm for each SAD system. Further, we compared the proposed SAD system a state-of-the art SAD that consists of a neural network trained on multiple feature streams~\citet{van2013robust}.  This system was developed at USC for DARPA RATS data. We can see from Table~\ref{table_sad} that proposed SAD is competitive with respect to lower false alarm rate and lower miss rate unlike USC-supervised-NN method that had significant false alarm.

The parameters of proposed diarization system was shown in Table~\ref{table_params_diar}. The results of diarization system were given in Table~\ref{table_results_der}. It is important to note that MFCC features from all the seven streams were used in HMM-based diarization for comparing its performance with bottleneck features. We could see that bottleneck feature captured useful statistics of multi-stream audio data that resulted in better accuracy using informed HMM-based diarization system. 

We extracted 13-dimensional MFCC features from each of the seven streams of the PLTL session. After concatenating the features from each stream we get a feature super-vector of dimensions 91 (=13*7). After splicing the feature super-vectors with five past and future frames (see Figure~\ref{fig_diar}), we get the final dimension of features as 1001 (=11*91). Spliced feature super-vector was fed to a stacked autoencoder for extracting the bottleneck features of dimension 21.  Stacked autoencoder with three hidden layers was chosen where the middle hidden layer acted as bottleneck layer. The bottleneck features were fed to the informed HMM-based diarization system. We used the Oracle SAD in the proposed system to validate the accuracy of HMM-based joint segmentation and clustering. However, we performed another case-study by formulating non-speech as an additional HMM state. We compared the diarization accuracy of bottleneck features (derived from raw MFCC features form each of the seven steams) and raw acoustic features (13-MFCC from each of the seven streams). Thus, the concatenation of MFCC features from multi-stream was done in both cases ensuring that it was a fair comparison between two approaches (raw features and bottleneck).

Table~\ref{table_results_der} showed the diarization accuracy in various cases. The "NO SAD" case refers to not using any SAD labels and modeling non-speech as an additional HMM state. We knew that the non-speech has several distinct varieties, such as silences (with extreme noise of varied types), overlapped speech~\emph{etc.}. This made the diarization, a challenging task without SAD labels. It led to degradation in diarization accuracy (see Table~\ref{table_results_der}). We could see that the bottleneck features combined with HMM was robust with respect to change in minimum duration constraints and to some extent is robust to absence of SAD labels. The state-of-the-art LIUM baseline~\citet{meignier2010lium} was borrowed from our earlier work for comparison~\citet{dubey2016interspeech}. We could see an absolute improvement of approximately 27\% in terms of DER over the baseline
LIUM system and approximately 12\% improvement was due to bottleneck features instead of using raw MFCC features (Oracle SAD, one second time-constraint).

Since the proposed dominance score, $DS$ (see Section~\ref{sec:dominance}) was derived using unsupervised acoustic analysis, we used Pearson's correlation between ground-truth dominance rating ($D_{rate}$) and proposed dominance score ($DS$). The correlation between ground-truth $D_{rate}$ and proposed $DS$ was 0.8748 for Eval-Set-3 (70 minutes). The high correlation value validates the efficacy of proposed dominance score, $DS$, for characterizing dominance in PLTL sessions. 

We computed pitch contour for each 25ms frame with 10ms skip-rate. For each speaker segment, we incorporated two-second of past and future speech and did the processing using non-overlapping super-segments of two-second duration. The pitch contour of each speaker segment was processed separately for possibility of detecting emphasis or question-inflection. We computed the speech energy based on wavelet packet decomposition as discussed in section~\ref{sec:wpd}. We computed the correlation between ground-truth mid-point and algorithm detection point as figure of merit for both emphasis and question-inflection detection. We also computed root mean squared error using ground-truth mid-point and detected points. Thus, we have two figures of merits namely correlation and root mean squared error for both, question-inflection and emphasis detection.

Table~\ref{table_results_pitch} showed the results on respective evaluation set. In addition, we did the detection error trade-off (DET)  analysis for question-inflection as depicted in Figure~\ref{fig_eer}. For getting DET curve we did not use the oracle diarization. We estimated the pitch contour for complete signal and performed the mean and variance normalization over non-overlapping super-segments of duration two-seconds. Thus, the newly generated normalized pitch contour was used for detecting the question-inflection by choosing a threshold. The values that were higher than the threshold corresponds to question-inflection. All frames between start-time and end-time of ground-truth question boundary were used as question-inflection points for the DET analysis. For each of the chosen threshold, we get a probability of false detection, $Pfa$ and probability of missing a true question-inflection, $Pmiss$. Figure~\ref{fig_eer} shows the detection error trade-off (DET) curve and the equal error rate (EER), where $Pmiss= Pfa$. The EER value comes out to be 12.31\%. The DET and EER analysis supported the hypothesis that pitch was a robust feature for detecting the question-inflection. Table~\ref{table_results_pitch} shows the results on evaluation datasets. 
\section{Conclusions}
This paper is a first step towards leveraging speech technology for extracting behavioral characteristics in small-group conversations such as PLTL sessions. Proposed methods were evaluated on CRSS-PLTL corpus. However, these algorithms can be extended to other similar applications such as small-group meetings/conversations. 

We established the CRSS-PLTL corpus that contains audio recording of five Peer-Led Team Learning (PLTL) team. Each team has six to eight students and a peer leader. This corpus provides an opportunity for researching speaker diarization and behavioral signal processing for multi-stream data collected in naturalistic scenarios. 

We used robust front-end for speech activity detection (SAD) and speaker diarization. Speech segments from all speaker were later processed with behavioral speech processing block that incorporate several acoustic analyses. Speech algorithms extract features capturing the behavioral characteristics such as participation, dominance, emphasis, curiosity and engagement. Results obtained on CRSS-PLTL corpus using proposed techniques are encouraging and motivate use of behavioral speech processing for understanding practical problems in education, human-to-human communication and small-group conversations. 

\section*{Acknowledgments}
Authors would like to thank anonymous reviewers and editors for helpful comments and suggestions that helped in improving the quality of this paper. 

Authors would like to thank Professor Shrikanth S. Narayanan, Maarten Van Segbroeck, and Andreas Tsiartas from University of Southern California, USA for providing the software of supervised Neural Network speech activity detection developed for DARPA RATS data.

We would like to thank Ali Ziaei (PullString Inc.), Lakshmish Kaushik and other CRSS colleagues for helpful suggestions and discussions.

This project was funded in part by AFRL under contract
FA8750-15-1-0205 and partially by the University of Texas at Dallas
from the Distinguished University Chair in Telecommunications
Engineering held by J. H. L. Hansen.

\section*{References}
%\bibliographystyle{model1-num-names}
%model5-names
%\bibliographystyle{model1-num-names}
%model1-num-names.bst
\nocite{*}
\bibliographystyle{model4-names}
\bibliography{ref_jp5}
\end{document}